\documentclass[11pt]{article}
\usepackage[a4paper, total={6in, 8in}]{geometry}

\usepackage{hyperref}
\usepackage{%
	amsmath,%
	amsthm,%
	bbm,%
	float,%
	etoolbox,%
	mathrsfs,%
	mathtools,%
	multirow,%
	pgf,%
	pgfplots,%
	subfigure,%
	tikz,%
}

\usepackage{amsfonts}
\usepackage{amssymb}

\usepackage{algorithmic}
\usepackage{xcolor}
\usepackage{thmtools,thm-restate}

\usepackage{todonotes}

\usepackage{bm}

\usepackage{graphicx}
\usepackage{tabularx}

\usepgflibrary{shapes}
\usetikzlibrary{%
	arrows,%
	patterns,%
	positioning,%
	shapes.callouts,%
	shapes.symbols,%
}
\usetikzlibrary{external}
\usepackage[ruled,vlined]{algorithm2e}
\usepackage{breqn}

\newcommand\numeq[1]%
  {\stackrel{\scriptscriptstyle(\mkern-1.5mu#1\mkern-1.5mu)}{=}}

\theoremstyle{plain}
\newtheorem{thm}{Theorem}
\newtheorem{lem}{Lemma}

\theoremstyle{definition}
\newtheorem{defn}{Definition}

\newtheorem{rem}{Remark}
\newcommand{\Nats}{\mathbb{N}}

\newtheoremstyle{case}{}{}{}{}{}{:}{ }{}
\theoremstyle{case}

\newcommand{\blue}[1]{\textcolor{blue}{#1}}

\newcommand{\norm}[1]{\left\lVert#1\right\rVert}

\newcommand{\s}{\mathcal{S}}
\newcommand{\R}{\mathbb{R}}

\newcommand{\E}{\mathbb{E}}
\newcommand{\Z}{\mathbb{Z}}

\newcommand{\mb}[1]{\mathbb{#1}}

\newcommand{\brac}[1]{\left(#1\right)}
\newcommand{\cbrac}[1]{\left\{#1\right\}}
\newcommand{\sbrac}[1]{\left[#1\right]}

\newcommand{\floor}[1]{\lfloor #1 \rfloor}

\newcommand{\TV}{{\scriptscriptstyle T \mspace{-1mu} V}}
\newcommand{\mix}{{\scriptscriptstyle m \mspace{-1mu} i \mspace{-1mu} x}}
\newcommand{\MW}{{\scriptscriptstyle M \mspace{-1mu} W}}

\newcommand{\mf}[1]{\mathbf{#1}}

\DeclareMathOperator*{\argmax}{arg\,max}

\title{Optimal Scheduling in a Quantum Switch
}

\author{Sanidhay Bhambay$^1$ \and  Thirupathaiah Vasantam$^2$ \and Neil Walton$^1$}
\date{%
    $^1$Durham University Business School\\%
    $^2$Department of Computer Science, Durham University   
}

\begin{document}
\maketitle

\begin{abstract}
 With a growing number of quantum networks in operation, there is a pressing need for performance analysis of quantum switching technologies.
  A quantum switch establishes, distributes, and maintains entanglements across a network. 
  In contrast to a classical switching fabric, a quantum switch is a two-sided queueing network. The switch generates Link-Level Entanglements (LLEs), which are then fused to process the network's entanglement requests.
  Our proof techniques analyse a two-time scale separation phenomenon at the fluid scale for a general switch topology. This allows us to demonstrate that the optimal fluid dynamics are given by a scheduling algorithm that solves a certain average reward Markov Decision Process.
\end{abstract}

\section{Introduction}
Quantum networking is required for multiple applications like Quantum Key Distribution (QKD), quantum communications, clock synchronization, distributed quantum computation, and quantum sensing \cite{wehner_qinternet,calefi_qi,HK_LO_qinternet}.
Recent deployments demonstrate that the Quantum Internet is no longer a theoretical concept: it is a fast-growing reality. Various large-scale QKD networks have been successfully assembled and connected in recent years. 
In China, a nationwide QKD network spans over 1,120 km, connecting major cities like Beijing and Shanghai~\cite{yin2020entanglement}. Europe has established its own quantum communication network~\cite{europeancommission2021}, and London's Finance Districts host the first commercial infrastructure for quantum-secure communications~\cite{toshiba2021quantum}.  
Further metropolitan demonstrators exist in China, the Netherlands, and the US
\cite{day2024entangled,liu2024creation,castelvecchi2024quantum}, and with China's satellite QKD network now connecting to Europe via Austria, there is a growing impetus to expand and connect these networks.
Quantum networks use switches that enable entanglement swapping.
Quantum switching technology is becoming increasingly important, with quantum switches becoming the core components of a quantum network. Similar to the current Internet, which uses high-speed switching technologies, we need to develop efficient programmable quantum switching algorithms to deploy quantum networks \cite{armstrong2012programmable}.

An important application for quantum switches is in distributed quantum computation. Here several small quantum computers are connected via a quantum network \cite{calefi_qdc,calefi_qi}. This is anticipated to be a requirement for solving complex computational problems which require millions of qubits \cite{quantum_chemistry}. (We note that qubits are basic units of quantum information that represent information embedded in some properties of atoms like polarizations of photons or spins of electrons, etc \cite{HK_LO_qinternet}). However, the current technology only allows us to have hundreds of qubits in a single quantum computer due to the lack of technology required to build large-scale cooling systems \cite{Cooling_systems}. Hence, we need to utilize several quantum computers together to make them communicate via a quantum network to solve realistic problems. 
The practical implementation of Quantum computing is at an early stage.\footnote{A current road map to Quantum computing is given here: https://quantumai.google/}
There are competing paradigms for quantum computation: circuit-based \cite{nielsen2001quantum}, annealed \cite{rajak2023quantum}, measurement-based \cite{briegel2009measurement}, and Fusion-based \cite{bartolucci2023fusion}. Several companies are pursuing Fusion-based technology due to its potential for implementation with photonic light at room temperatures. Here, switching technology has been essential \cite{bartolucci2021switch} to build an efficient quantum network for connecting computing nodes, and thus, there is an interest in the performance evaluation of quantum switches. Regardless of the technology used, there is a strong belief that any large-scale quantum infrastructure will require network infrastructure in much the way high-performance computing does. Through this, again, we require entanglement distribution through photonic switch technology.

For applications in the quantum internet, the primary requirement is to share entanglements, which are multi-qubit quantum states with correlated qubits, with distributed users in such a way that each user holds one qubit of the entanglement. For example, quantum communication can be achieved between two nodes by quantum teleportation once a pre-existing entanglement is shared between them. To distribute entanglements to remote users, one could create entanglements at a single location using entanglement sources and then transfer the qubits of entanglements to desired users. However, the transmission of qubits through communication channels has significant challenges. Qubits decay over time when they are transferred across channels as they suffer from noise in communication channels, and intermediate repeaters cannot restore or boost signals as in classical communication due to the No-cloning Theorem \cite{calefi_qi}. If qubits are photons, then the successful transmission of photons in fiber-optic channels between two nodes decays exponentially with the length of the channel; they can only travel up to a distance of 100km in real systems \cite{HK_LO_qinternet}. Therefore, it is challenging to distribute entanglements to distributed users when the distance between them is large.

To overcome the challenges of distributing entanglements to remote users, quantum repeaters or switches apply entanglement swapping operations. Here, repeaters are connected to only two neighboring nodes, while switches are connected to multiple neighboring nodes and have switching logics. A quantum network is a network of quantum devices like repeaters, switches, and users that run applications. The length of individual links are sufficiently small enough to create entanglements between neighbouring nodes with high success probability by direct transmission of photons across links, such entanglements are called Link-Level Entanglements (LLEs). The switches fuse LLEs of neighboring links using quantum measurement operations to produce entanglements distributed to neighboring nodes. In a quantum network, LLEs of a link could be used to distribute entanglements for different applications or user-groups. Hence, effective routing and scheduling algorithms should be designed to resolve contention and distribute entanglements required by diverse applications. In order to produce entanglements at higher rates, links produce LLEs continuously and switches can fuse LLEs produced at different time instants. Unused qubits are stored in quantum memories at end nodes of the link. When LLEs are stored in this way, they may decohere or expire after a finite amount of time due to noise in quantum devices. For further information, We recommend the reader \cite{HK_LO_qinternet,calefi_qi} for a detailed discussion and introduction to quantum networking.



If we focus on the performance of a single quantum switch. From the description above, we see that a quantum switch is a two-sided queueing system: On the one side—similar to a ``traditional" cross-bar queue switch  -- requests arrive requiring access to LLEs of a group of links. However, on the other side, a separate random process generates LLEs that are either consumed immediately or stored in quantum memories for future usage. An LLE may decohere over time. Further, the quantum measurements-- Bell-state measurements to fuse two LLEs and Greenberger–Horne–Zeilinger (GHZ) measurements to fuse more than two LLEs-- succeed only with certain probabilities.  
The decision to schedule requests over a given time period creates non-trivial interdependence between the stochastic processes of links and requests.

Traditional switch fabric in the current Internet consists of a bipartite matching between input and output ports \cite{ramaswami2009optical}. Here, there are constraints on service in that only one input can be matched to one output, and the scheduling decision made in one round does not impact the availability of future scheduling choices. Here, the set of arrival rates that can be stabilized is well understood.  In this case, the set of arrival rates that can be stabilized is well-understood. Essentially, it is the convex combination of the set of scheduling decisions \cite{georgiadis2006resource}. 
The MaxWeight algorithm is an indispensable for scheduling theory in queueing networks. It can stabilize a network for all arrival patterns without prescriptive knowledge of arrival statistics, which is why such algorithms are called throughput optimal.
The MaxWeight algorithm was first introduced in the context of wireless ad-hoc networks \cite{tassiulas1990stability} and then developed for Internet Switch design by \cite{mckeown1999achieving}. This led to the design of the iSlip algorithm, which was then commercially deployed by Cisco \cite{mckeown1999islip}. 

There is considerable interest in quantum switches by computer scientists \cite{zubeldia2022matching,Longbo}, electrical engineers \cite{valls2024brief,promponas2023full}, physicists \cite{collins2005quantum} and industry\footnote{See also: \url{https://www.psiquantum.com/blueprint} and \url{https://www.youtube.com/watch?v=U5pRnK7dGcI}} \cite{bartolucci2021switch,mandil2023quantum,palacios2018introduction} to understand the design and control of these systems. 
However, when we turn to a quantum switch as a two-sided queueing model, it is no longer clear what the capacity region is,
given that LLEs can survive from one-time slot to the next. 
In this paper,
we first characterize the capacity of a general quantum switch topology. Previous works have studied only simplified quantum switch models. Nain et al. \cite{Nain_switch} find a quantum switch model's capacity region with only links and no request queues. Quantum switches with two-sided queues with one-time slot and infinite lifetimes for qubits of LLEs are studied in \cite{Thiru_switch,10229003} and \cite{Wenhan_switch}, respectively. In recent work \cite{zubeldia2022matching}, a quantum switch model similar to ours is studied; however, their analysis is limited to specific $Y$ and $W$ shaped topologies for two types of requests and three links. The analysis that we complete here, extending the analysis to a general two-sided switch model, allowing general LLE structure, characterizing the capacity region, and demonstrating throughput optimality, are identified to be a significant challenge in prior works \cite{Nain_switch,Wenhan_switch,zubeldia2022matching}.  

This paper addresses the question: Can we maintain optimal throughput in a generic Quantum switch? We show that the answer is yes. This is due to the time-scale separation that can occur between the request queues and the link-level entanglements when the switch is congested. 
Timescale separation has a long interest in the performance analysis of queueing systems \cite{hunt1994large,10.1214/14-AAP1057,yasodharan2022large,10.1145/3152042.3152052}. Timescale separation is first investigated for the MaxWeight policy in a quantum switch with $Y$ and $W$ topologies \cite{zubeldia2022matching}.  
Under a general topology, we find that asymptotic decoupling occurs between the request and entanglement queues when a switch becomes congested with requests. Surprisingly, we find the MaxWeight algorithm is not throughput optimal. This is because it is no longer the optimal scheme to minimize instantaneous drift (See Figure~\ref{Fig:lyapunov_drift}).
Instead, we find optimal scheduling requires optimizing an average reward Markov Decision Process (MDP), which only depends on requests through the queue size and the dynamics of the entanglement generation process. Thus, if we can learn the switch's internals sufficiently well to solve this MDP, we can construct throughput optimal scheduling algorithms for the quantum switch. Through this, our paper develops several new analytical techniques on two-sided queueing models that can be used to make further advances in quantum networking.












\subsection{Our contributions}
We design an asymptotically optimal scheduling scheme to tackle the problem of scheduling LLEs based on entanglement requests in a quantum switch. 
In this context, our contributions are:
\begin{enumerate}
    \item We analyse a quantum switch with a general graph topology. Due to its technical difficulty, this setup is not examined in prior works. We provide the first characterization of the capacity region of these switches. By focusing on this more complex and generalized topology, our work addresses a broader and more challenging problem of scheduling LLEs in quantum switches instead of the more restricted models considered in earlier works.
    \item While the MaxWeight policy is known for achieving optimal throughput classical switches, it turns out that this is not the case for quantum switches. We provide a simple counter-example to the through-put optimality of MaxWeight.
    \item For the scheduling of LLEs, this work proposed a novel, asymptotically optimal scheduling scheme called the \textit{Average Reward Entanglement} (ARE) policy. The ARE policy solves an average reward MDP and is designed to maximize the efficient utilization of LLEs to serve incoming requests for entanglements. This is the first work to propose an asymptotically optimal scheduling policy for general quantum switches.
    \item To establish the optimality of the ARE policy, several technical contributions are made. As the quantum switch becomes congested, the LLEs will evolve on a faster timescale compared to the incoming entanglement requests, introducing a two-time scale separation. A novel approach has been developed in this work to establish the fluid limit using this time-scale separation. In this approach, time is divided into equal segments, with the length of each segment carefully chosen to ensure that the faster process reaches its steady state within each segment before the request process undergoes significant change in its state. The faster process's steady-state behavior then governs the slow process's dynamics. 
    \item Our analysis shows that the ARE policy achieves the optimal drift. We prove the asymptotic throughput optimality of ARE. Thus, the ARE policy provides more efficient scheduling and outperforms the traditional MaxWeight policy for quantum switches. 
\end{enumerate}

At a methodological level, the interplay between time-scale separation and Markov decision processes in optimal scheduling of matching queues is intriguing and potentially of interest to the performance evaluation of general matching systems outside of quantum networking problems. Our policy demonstrated novel methods for characterizing and optimizing capacity for the quantum system that are significantly more general than the current state-of-the-art in quantum switches. These results resolve several problems that were previously open on quantum switches \cite{Nain_switch,Wenhan_switch,zubeldia2022matching}.

Quantum technologies will likely be one of the most important areas for the 
modeling and performance evaluation of computing systems in the next decade. In the thirty years since Quantum communication was first initiated \cite{bennett1993teleporting},
there have been many changes: the growth of the Internet, the web, mobile communications, social networks, platforms, data centers, machine learning, and AI. 
Given this, it is reasonable that the attention of the performance evaluation community has been elsewhere.\footnote{To the best of our knowledge, one SIGMETRICS conference paper has been published on Quantum communication since inception over  thirty years ago \cite{nain2020analysis}: \url{https://dl.acm.org/action/doSearch?AllField=Quantum\&expand=all\&ConceptID=119981}}
Thus, aside from the specific technical contributions outlined above, one aim of this paper is a \emph{call-to-action} emphasizing the importance and timeliness of quantum communication. We firmly believe there is great potential in the performance evaluation community to contribute to the measurement and methodology that will underpin these emerging technologies.

\subsection{Organization}
In Section~\ref{sec:system_model}, we define our primary model. This section introduces basic mathematical notation, the topology of a quantum switch, and a description of its dynamics. We also recall several results on the mixing times of Markov chains.
In Section~\ref{sec:policies}, we discuss scheduling policies, demonstrate that MaxWeight is suboptimal, and define the capacity region of these Markov chains as a function of the stationary LLE process. We further discuss time-scale separation and use this to give an optimal policy for these systems. Our theoretical findings are presented in Section~\ref{sec:main_results}. These findings consist of a characterization of the fluid limit of a quantum switch. The proof of this limit requires a timescale separation analysis of these systems. We demonstrate the optimality of the fluid limit of our policy, and we prove the asymptotic optimality of our policy.
In Section~\ref{sec:proofs}, we prove the main proofs of the paper; however, due to space constraints, we must postpone several important arguments to the appendix.

\section{System Model and Dynamics}
\label{sec:system_model}
This section provides a detailed discussion of the system model and dynamics of the quantum switch analyzed in this paper. The key components include the switch topology and the switch dynamics.

\subsection{Basic Notation}
We denote  the set of non-negative integers as $\Z_+$ and  $[M]$ corresponds to $\{1,\ldots, M\}$. We use $\bar{\mathbb{R}}$ to denote the set of extended reals, i.e., $\R\cup \{\infty\}$. For $x,y \in \mathbb R$, $x\vee y = \max\{x,y\}$ and $x\wedge y=\min\{x,y \}$. For $i \in [M]$, $\bm e_i$ is the $i$-th unit vector in $\mathbb R^M$; $\bm 1$ is the vector of all ones; $\bm 0$ is the zero vector. For vectors $\bm u, \bm v \in \mathbb R^M $, we write $\bm u \leq \bm v$ if $u_i \leq v_i$ for $i\in[M]$. For a set $\mathcal{M}$, we use $|\mathcal{M}|$ to denote its cardinality. We use $\bm a \cdot \bm b$ to denote the dot product and $\bm a \bm b$ to represent the element-wise multiplication between two vectors of the same dimension. We use the notation $[y]_+=\max(y,0)$ and $[y]_+^a=\max(y,a)$.

\subsection{Switch Topology}
\label{subsec:topology}
We investigate a quantum switch system where multiple links connected to the switch generate LLEs or Bell-pairs. The system needs to accommodate requests requiring end-to-end entangled states formed by merging LLEs of various links through entanglement swapping. An $n$-qubit entanglement can be formed by applying GHZ measurements (if $n\geq 3$) or Bell-state measurements (if $n=2$) on a group of LLEs such that each LLE belongs to a different link. In this context, we consider $L$ links and $R$ types of requests. The switch can fulfill each request type by combining these LLEs. Requests of different types might require an individual link's LLEs, in which case there is a competition among these requests to use LLEs of the link.

\subsubsection{Requests and Links} We consider a discrete-time system where, at the beginning of each time slot, requests of $R$ types arrive. We denote $\mathcal{R}$ as the set of $R$ request types and $\mathcal{L}$ as the set of $L$ links. The switch stores requests of each type in their corresponding infinite-capacity queue and processes them on a First-Come-First-Served (FCFS) basis. 
A type $r\in \mathcal{R}$ request is considered served when the switch establishes an $n_r$-qubit end-to-end entanglement among a set of links denoted by $\mathcal L_r = \{l_{r,1}, \dots, l_{r,n_r}\} \subset \mathcal{L}$.
Here, $\mathcal L_r$ represents the set of links whose LLEs are necessary to fulfill a type $r$ request. A type $r$ request is successfully served if the measurement operation applied on LLEs of the links in $\mathcal L_r$ is successful. Note that LLEs used in measurement operations are always discarded irrespective of the outcome of measurement operations. 

\begin{figure}[h!]
  \centering
  \includegraphics[width=10cm]{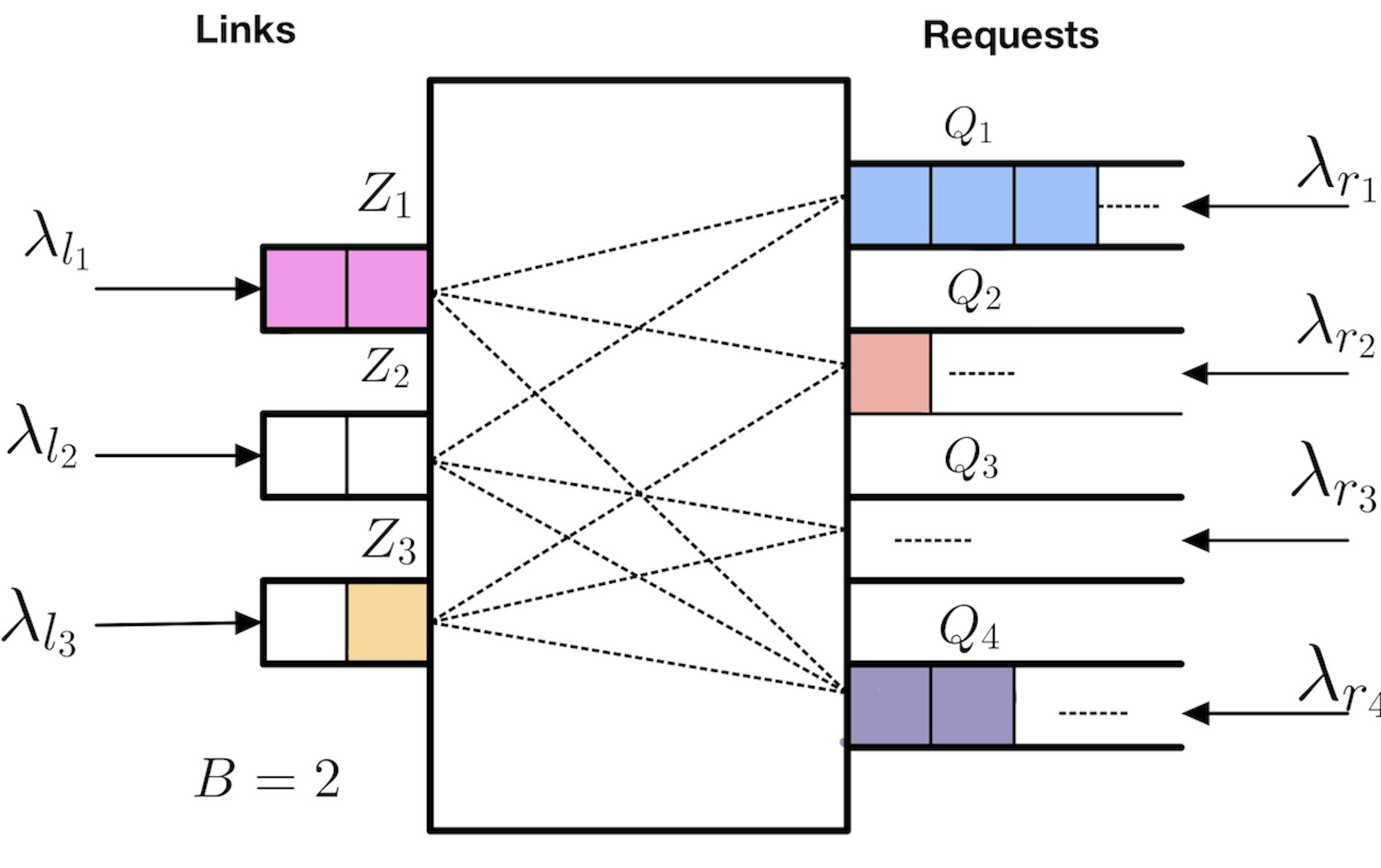}
\caption{A quantum switch with four request types and three links is considered. For this switch, $B = 2$. Requests arrive with a mean rate of $ \lambda_{r_i} $ for $ i \in [4] $, while LLEs are generated at a mean rate of $ \lambda_{l_i} $ for $i \in [3] $. }
\label{Fig:sm_qs}
\end{figure}

Since we consider a two-sided matching system, we will typically consider requests on the \emph{right-hand side} and the links on the \emph{left-hand side}. See Figure~\ref{Fig:sm_qs}. 
\subsubsection{Success Probabilities and Buffer Size.}
The success probability of applied measurement operations for a type $r$ request is denoted by $\gamma_r\in(0,1]$.
Furthermore, at the beginning of each time slot, the switch attempts to create new LLEs between each link $l$ and the switch for all $l \in \mathcal{L}$. 
We assume that the switch and the users have a finite storage capacity for LLEs, given by the buffer size $B$. In other words, each link in the switch can hold up to $B$ LLEs at any given time, beyond which any additional LLEs cannot be accommodated and are discarded. This assumption agrees with the fact that users and the switch will have a finite number of quantum memories in practical quantum systems, as quantum memories are expensive to use.\footnote{We note that from a theoretical perspective, infinite buffers are likely possible to analyze so long as LLEs have a strictly positive decoherence probability. However, our proof and mixing time bounds are considerably simpler with finite buffers, which is currently more realistic in Quantum repeaters.} On the other hand, requests can wait until served. In the current Noisy Intermediate-Scale Quantum (NISQ) computers era, a quantum computer has only hundreds of qubits and quantum devices and gates are noisy~\cite{NISQ1, NISQ2}. Qubits decohere easily because of depolarizing and dephasing noise as a result of their interactions with the environment. Hence qubits have finite lifetimes (coherence times) in real systems; these lifetimes vary depending on the type of the technology used to prepare qubits \cite{quantum_memory}.
 As in \cite{Fittipaldi}, we make an assumption that at the end of each time slot, each unused LLE of link $l$ has an independent probability $d_l\in(0,1]$ of being discarded; this models the scenario where the storage-and-retrieval efficiency of a quantum memory in a time slot is $e^{-\frac{\Delta t}{\nu}}$ where $\nu$ is the lifetime of a qubit and $\Delta t$ is the length of the time slot.

\subsubsection{Schedules.} We use discrete-time scheduling policies for the switch topology described above to create entanglements for the different arriving requests. 
We focus on scheduling requests before discussing their implications on the LLEs. We let $\bm n = (n_r: r \in \mathcal{R})\in \mathcal N \subset \mathbb Z_+^R$ denote the number of requests that can be processed in a time slot, and here $\mathcal N$ is the set of requests that can be scheduled in a given time slot.
We assume that the set $\mathcal N$ is monotone in that if $\bm n \in \mathcal N$ then $\bm n' \in \mathcal N$ for all $\bm n'\in \mathbb Z_+^R$ satisfying $\bm n'\leq \bm n$. We note that this assumption will ensure that our policy and the MaxWeight policy is work conserving.

The number of requests scheduled has implications for the number of LLEs queued.
A request of type $r$ requires an LLE from each of its links: 
$$\bm \psi_r := \sum_{l \in \mathcal L_r}\bm e_l \,.$$
Since we allow for more than one request to be processed per time slot, for $\bm n = ( n_r \in \mathbb Z_+)$  and $\mathcal N$, we can respectively define a schedule and the set of schedules as follows
\begin{equation}
\bm \sigma (\bm n) := \bm n^\top \bm \psi = \sum_{r\in \mathcal{R}} n_r \bm \psi_r   \qquad \text{and}\qquad  \mathcal S := \left\{ (\bm \sigma(\bm n),\bm n) :  \bm n \in \mathcal N\right\}\, . \label{eq:sigma(n)}    
\end{equation}

\subsection{Quantum Switch dynamics}
We now describe the dynamics of the quantum switch discussed in Section \ref{subsec:topology}.  Here we consider a discrete-time model.

\subsubsection{Order of events.} Over one time-slot,
the following sequence of events occurs. First, new LLEs are generated between each link and the switch.
Next, the number of requests of each type for service is decided. Then, an LLE schedule is selected. 
A number of these scheduled requests will fail.
These requests will not be removed, but the LLE will have decoherence and thus will be removed.
Next, toward the end of the time slot, new arrivals for requests occur.
Finally, at the end of the time slot, a number of leftover LLEs will decohere.

\subsubsection{Arrivals.} Type $r$ request arrive according to an exogenous arrival process $A_r(\cdot)$ whose increments $a_r(t) := A_r(t)-A_r(t-1)$ are i.i.d. with mean $\lambda_r \in \R_+$ and are positive and bounded. 
The arrivals to the request queues are assumed to be independent of each other. 
Furthermore, for each $l \in \mathcal{L}$, LLEs are generated between link $l$ and the switch according to the process $A_l(\cdot)$. Again the increments $a_l(t) := A_l(t)-A_l(t-1)$ are i.i.d. with mean $\lambda_l \in \R_+$ and are positive and bounded.  



\subsubsection{Queue Lengths.} We define the queue-length vector of requests at the beginning of time slot
$t$ as
$\mf Q(t)=(Q_r(t),r \in \mathcal{R}),$ 
where $Q_r(t)$ denotes
the queue length of the type $r$ request. We also define the LLE-vector at the beginning of time slot
$t$ as $\mf{Z}(t)=(Z_{l}(t),l \in \mathcal{L}),$
where $Z_l(t)$ represents the total number of link $l$ LLEs at the beginning of time slot $t$. 
We define the joint process $\bm X(t) := (\bm Q(t),\bm Z(t))$
with
$\bm X =(\bm X(t):t\in \Z_+)$. 
For the policies considered in this paper, $\bm X$ will be a Markov process.
When the system is positive recurrent, we use $\bm X(\infty) = \brac{\mf Q(\infty),\mf Z(\infty)}$
to denote the steady-state values. 
Moreover, we define the fluid scaled processes $\bar{\bm X}^c=\brac{\bar{\bm X}^c(t): t\in \Z_+}$ where $\bar{\bm X}^c(t)=\brac{\bar{\bm Q}^c(t),\bar{\bm Z}^c(t)}$  such that
\begin{equation*}
\bar{\bm Q}^c(t)=\frac{\mf Q(ct)}{c}, \qquad  \bar{\bm Z}^c(t)=\frac{\mf Z(ct)}{c}, \qquad\quad \forall t\in \Z_+, \ c\in \Nats.
\end{equation*}
\subsubsection{Schedules and Departures.} We decide the number of requests to schedule in each time slot. 
We let $\bm n(t)= (n_r(t) \in \mathbb Z_+ : r\in \mathcal{R}) \in \mathcal N$ be the number of requests of each type attempted at time $t$. We apply the abbreviation $\bm \sigma(t) := \bm \sigma(\bm n(t))$ to give the number of LLEs scheduled, where $\bm \sigma(\bm n)$ is defined in \eqref{eq:sigma(n)}.
The number of requests and scheduled LLEs is chosen to be less than the number of existing requests and LLEs. 




A number of attempts to serve requests will fail, we define $\hat{\bm n}(t)= (\hat n_r(t) : r \in \mathcal{R})$ to be the number of successful requests:
\begin{equation}
\label{eqn:success_requests}
 \hat n_r (t) \sim \textsc{Binomial}(n_r(t) , \gamma_r).  
\end{equation}
We can define the cumulative number of departures of requests and links, respectively, with 
\[
D_r(t) = \sum_{s =1}^t \hat n_r(s)\qquad 
\text{and}\qquad D_l(t) = \sum_{s =1}^t \sigma_l(s)\,.
\]
Furthermore, we assume that $D_r(0)=0$ for $r\in \mathcal{R}$ and $D_l(0)=0$ for $r\in \mathcal{L}$.

\subsubsection{Queue Dynamics} \label{sec:Qdynamics}
The request queue process $\mf Q=(\mf Q(t):t\in \Z_+)$ evolves according to following equations 
\begin{align}
\label{eqn:request_evolution}
Q_r(t+1)=[Q_r(t) - \hat {n}_r(t)]_+ + {a}_r(t) \, ,
\end{align}
for $t\in \Z_+$ and for $r\in\mathcal{R}$. 
We let $Y_l(t)$ be the state of the link after arrivals but not accounting for decoherence. Note that $Z_l(t)$ is then a binomial RV given $Y_l(t)$:
\begin{align*}
    Y_l(t) 
    &
        = [[ Z_l(t-1) + a_l(t)]_+^B - \sigma_l(t)]_+ \, ,
    \quad \text{and} \quad 
    Z_l(t) 
    \sim \textsc{Binomial}(Y_l(t),d_l)\, .
\end{align*}

\begin{rem}[General Decoherence Times]
    Above, we consider i.i.d. decoherence probabilities $d_l$. However, it is possible to consider general lifetime distributions here. For this, we can consider the age distribution of each LLE in the buffer. Then we let $d_l(a)$ be the probability of decoherence in the next timestep of an LLE of age $a$. We must store all ages to give a Markov description of our switch. In other words, $Z_l(t)$ will be a vector of the ages of all LLEs at the link. 

\end{rem}

\begin{rem}[General Matching Systems]
The specific dynamics of $\bm Z$ described above are not critical to our results. We require $\bm Z$ process mixes once we condition on $\bm Q(t)$ and that the transitions in $\bm Z(t)$ are Markovian.
So we can allow for more general matching processes. 
For example, we could model a pool of vehicles in a ride-sharing system or a call centre with staff arrivals and departures.
\end{rem}

\subsubsection{Mixing Properties.}
For two (discrete) probability distributions $\bm \mu$ and $\bm \mu'$ and for a Markov chain transition matrix $P$, the total variation distance and the coefficient of ergodicity are, respectively:
\[
\| \bm \mu - \bm \mu' \|_{\TV} \,  = \frac{1}{2} \sum_{z \in \mathcal Z} |\mu_z - \mu'_z |
\quad \text{and} \quad \rho( P) = \sup_{\substack{\bm v} } \left\{ \| \bm v P \|_{\TV} :  \bm 1^\top \bm v = 0, \,\| \bm v \|_{\TV}=1 \right\}\, .
\]
%
We define $P^{\bm Q}$ to be the transition matrix defined by
\[
P_{zz'}^{\bm Q} = \mathbb P ( \bm Z(t+1) = \bm z' | \bm Z(t) = \bm z, \bm Q(t) = \bm Q ), \qquad z,z' \in \mathcal Z\, .
\] 

We can prove fast mixing occurs regardless of the state of $\bm Q$:
\begin{restatable}{lem}{lemrho}\label{lem:lemrho}
   There exists a constant $\rho$ such that for all $\bm Q$, 
    \[
        \rho(P^{\bm Q}) \leq \rho <1\, .
    \]
\end{restatable}
The Lemma holds essentially since the decoherence of LLEs leads to fast mixing of the $\bm Z(t)$ process.
From a theoretical perspective, it follows from Dobrushin's Lemma on the coefficient of ergodicity. This as well as some standard facts about $\| \cdot \|_{\TV}$ and $\rho(\cdot)$ are given in Appendix \ref{sec:mix}.

\section{An Optimal Scheduling Scheme}
\label{sec:policies}
The scheduling policy aims to determine how incoming requests are processed by selecting from the available schedules in $\mathcal{S}$. The scheduling policy establishes entanglements based on the switch's available LLEs at each link.

The key challenge is to balance the immediate use of LLEs to satisfy current requests against the potential benefits of preserving LLEs for future use. Traditional switch scheduling schemes, such as the celebrated MaxWeight scheme, focus solely on maximizing the number of entanglement requests fulfilled in the current time slot, such a MaxWeight scheme has been shown to be throughtput optimal when LLEs have lifetimes of one time slot \cite{Thiru_switch}. However, we argue here that such an approach could lead to suboptimal outcomes over the long term when LLEs survive for more than one time slot, particularly in scenarios where immediate satisfaction of all requests depletes the LLEs, leaving fewer resources available for future time slots.

We demonstrate that the optimal scheduling scheme for serving entanglement requests, based on the availability of LLEs, can be effectively formulated as an Average Reward MDP. The key to this approach is the time-scale separation that naturally arises between the link-level process and the request process. 



\subsection{Scheduling Policies}
A scheduling policy, which we will typically denote by $\pi$, is a rule that determines the scheduling decision in $\s$ at time $t$ based on the past observed state of the system $(\bm X(s): s\leq t)$. 
We allow for the possibility of randomized schedules. With this in mind, we define $\langle \mathcal S \rangle$ to be the set of random variables with support on $\mathcal S$.
%
%

For quantum switches, one of the objectives of scheduling policies is to maximize the throughput --  the rate at which requests are successfully served. 
However, the scheduling policies considered in this work must not only maximize the throughput of request types but also effectively manage the allocation of LLEs. 
This approach is distinct from prior work on scheduling in switches.


We refer to a scheduling policy as Markovian if it depends only on the current state $\bm X(t)$. Thus, a policy is a function from the set of states to the set of (randomized) schedules $\bm \pi : \mathcal X \rightarrow  \langle\mathcal S \rangle $. 
For $\bm x \in \mathcal X$, we will apply the notation:
\begin{equation} \label{eq:pishorthand}
    \bm \pi(\bm x) = (\bm \sigma(\bm x) , \bm n(\bm x)) \qquad \text{and} \qquad p(\bm \sigma,\bm n | \bm x)\, ,
\end{equation}
where the random variables $\bm \sigma(\bm x) = (\sigma_l(\bm x) : l \in \mathcal L)$ give the number of LLEs scheduled and the random variables $\bm n(\bm x) = (n_r(\bm x) : r \in \mathcal R)$ gives the number of requests scheduled when in state $\bm x$, and where $p(\bm \sigma,\bm n | \bm x)$ is the probability of selecting schedule $(\bm \sigma,\bm n)$, when the switch is in state $\bm x$.

We say a scheduling policy is \emph{request-agnostic} or \emph{agnostic} if it does not have knowledge of the number of requests. Specifically, a request-agnostic policy is a function from the set of LLE states to the schedules, $\bm \pi : \mathcal Z \rightarrow \langle\mathcal S \rangle$.
We let $\mathcal A$ denote the set of request-agnostic policies.
Similar to \eqref{eq:pishorthand}, we define $\bm \pi(\bm z) = (\bm \sigma(\bm z) , \bm n(\bm z))$ and $p(\bm \sigma,\bm n | \bm z)$ where $\bm \sigma(\bm z)$ gives the LLEs scheduled and $\bm n(\bm z)$ gives the requests scheduled when the LLEs are in state $\bm z$ and where $p(\bm \sigma,\bm n | \bm z)$ is the probability of selecting schedule $(\bm \sigma,\bm n)$, when the LLEs are in state $\bm z$. For an agnostic policy, we assume scheduled LLEs are removed regardless of the state of the request queue.  

\subsection{Capacity Region}
The capacity region is the set of request rates for which the system can be stabilized. More formally, the capacity region, $\mathcal C$, is the set of arrival rates for requests $\bm \lambda \in \mathbb R_+^R$ for which there exists a policy where the queue size process $\bm X(t)$ is a positive recurrent Markov chain.

When server availability is independent of the system state, there are standard arguments where the capacity region can be characterized, see \cite{andrews2004scheduling}. However, this is not the case for our matching system; there is an interdependence between both sides of the switch.
Thus, the characterization of the capacity region below is new and non-standard due to how service on the righthand queues is interdependent with the evolution of the lefthand-side queues. 

\begin{restatable}{thm}{ThrmStabilityRegion}\label{thrm:stabilityRegion}
 A necessary condition for $\bm \lambda \in \mathcal C$ is that there exists a request agnostic policy $\bm \pi $ with LLE process having stationary distribution $\mu$ such that 
\begin{equation}\label{eq:Cbar}
\lambda_r \leq \sum_{\bm z\in \mathcal Z}\mu(\bm z)\sum_{(\bm{\sigma} ,\bm{n})\in\mathcal{S}}p(\bm{\sigma},\bm{n} | \bm z)   n_r \gamma_r  \, , \qquad \forall r  \in \mathcal R\, .    
\end{equation}   
A sufficient condition for $\bm \lambda \in \mathcal C$ is that there exists a request agnostic policy $\bm \pi$ with LLE process having stationary distribution $\mu$ such that 
\begin{equation}\label{eq:Ccirc}
\lambda_r <  \sum_{\bm z \in \mathcal Z}\mu(\bm z)\sum_{(\bm{\sigma} ,\bm{n})\in\mathcal{S}}p(\bm{\sigma},\bm{n} | \bm z)   n_r \gamma_r  \, , \qquad \forall r  \in \mathcal R \, .   
\end{equation}
\end{restatable}

Due to space constraints, the proof of Theorem \ref{thrm:stabilityRegion} is given in Section~\ref{append:stabilityregion} in the appendix.

The above result essentially characterizes the set of arrival rates that can be stabilized. Given the above result, we define $\bar{\mathcal C}$ to be the set of arrival rates $\bm \lambda$ such that \eqref{eq:Cbar} holds, and we define $\mathcal C^\circ$ to be the set of arrival rates such that \eqref{eq:Ccirc} holds. 
For $\epsilon >0 $, we also define $\mathcal C^\epsilon$ to be the set of $\bm \lambda $ such that
\begin{equation}\label{eq:Ceps}
\lambda_r + \epsilon \leq \sum_{\bm z\in \mathcal Z}\mu(\bm z)\sum_{(\bm{\sigma} ,\bm{n})\in\mathcal{S}}p(\bm{\sigma},\bm{n} | \bm z)   n_r \gamma_r  \, , \qquad \forall r  \in \mathcal R\, .    
\end{equation} 
From Theorem \ref{thrm:stabilityRegion}, we have $\mathcal C^\epsilon \subset \mathcal C^\circ \subset \mathcal C \subset \bar{\mathcal C}$. Also if $\bm \lambda \in \mathcal C^\circ$ then $\bm \lambda \in \mathcal C^\epsilon$ for some $\epsilon>0$.  We now provide the definitions of throughput optimality and asymptotic throughput optimality. 
\begin{defn}
Policy $\bm \pi$ is \emph{throughput optimal} if it's positive recurrent for all arrival rates $\lambda\in \mathcal C^\circ$. 
\end{defn}

\begin{defn}
A set of policies $\brac{\bm \pi^c}_{c\in\mathbb N}$ are \emph{asymptotically throughput optimal} if for all $\epsilon >0$ there exists a constant $c_\epsilon$ such that for all $c \geq c_\epsilon$ the policy $\pi^c$ is positive recurrent for all $\bm \lambda \in \mathcal C^\epsilon$.   
\end{defn}
 Informally stated, a policy is throughput optimal if it is stable for the maximum set of arrival rates; a set of policies is asymptotically throughput optimal if there is a policy that can stabilize arrival rates within an arbitrarily small error of the maximum set of arrival rates.

We note that the set of stabilizable policies depends on the set of stationary distributions of the lefthand queueing system. To optimize service, we must optimize over the stationary distribution of the LLEs. This provides one justification for optimizing an average reward MDP for the lefthand queues to achieve throughput optimality.

\subsection{MaxWeight Scheduling}
Perhaps the most widely studied policy in the network community is the MaxWeight policy \cite{tassiulas1990stability,mckeown1999achieving}. It selects the schedule that maximizes a weighted sum of queue lengths in each time slot.
\begin{defn}(MaxWeight):
At any time $t$, the MaxWeight scheduling policy $\pi^{\MW}$ selects a schedule $\bm n^{\MW}$ as follows:
\begin{equation}
\label{eqn:MW_schedule}
\bm n^{\MW}(t)\in \argmax_{\bm n: (\bm \sigma(\bm n),\bm n) \in \s}\sum_{r\in \mathcal{R}}Q_r(t)n_r.
\end{equation}
\end{defn}
 At any time $t$, the schedule $\bm{n}^{\MW}(t)$ chosen by the MaxWeight policy is dependent on the request queue lengths $\bm{Q}(t)$. Thus, the selected schedule can be expressed as $ \bm{n}^{\MW}\brac{\bm{Q}(t)}$. 

For a classical switch \cite{mckeown1999achieving}, the MaxWeight policy is known to be maximally stable in that it stabilizes the system if the average arrival rates $\bm \lambda=\brac{\lambda_r,r\in\mathcal{R}}$  lie within the capacity region of the system that is $\bm \lambda \in \langle \mathcal S\rangle$. However, as we have seen in Theorem \ref{thrm:stabilityRegion}, the set $\langle S \rangle$ is no longer the capacity region, $\mathcal C$, for a quantum switch and, as we now discuss, this leads to a drop in performance when directly applying MaxWeight. 

The traditional MaxWeight policy applies to models that do not consider any form of randomization in the number of available servers. Thus, our policy generalizes the traditional MaxWeight policy: all MaxWeight policies are a subset of our policy. There are extensions of MaxWeight that consider an independent number of servers, see for example \cite{georgiadis2006resource}. The consideration of servers as a Markov process in the context given here is new to the best of our knowledge. Under this extension, we could assume that we have queues associated with the number of available servers. However, given that the number of servers in the quantum setting is $O(1)$ (and bounded in our case), these extensions do not provide any additional structural properties. For example, the counterexample network considered below will still apply to these extensions.


\subsection{A Counter Example: MaxWeight is not Throughput Optimal.}
\label{subsec:counter_example}
There has been interest in implementing MaxWeight in quantum switches \cite{zubeldia2022matching,Thiru_switch,10229003}.
Here, we provide a counter-example that proves MaxWeight is not optimal throughput in a quantum switch when LLEs decohere after more than one timeslot. The logic of this counter-example is similar to the linear network counter-example of Boland and Massouli\'{e} \cite{bonald2001impact}.

Suppose that there are three links $l_1$, $l_2$, $l_3$, and three types of request: $r_1$ which requires one LLEs from link $l_1$ only; $r_2$ which requires an LLE from link $l_2$ only; and $r_3$ which requires an LLE from every link $l_1$, $l_2$, $l_3$. 
Suppose that one LLE arrives every three units of time at each link. Of these three time slots, the switch generates a link $l_1$ entanglement in the first time slot. In the second time slot, the switch generates a link $l_2$ entanglement.  In the third, the switch generates a link $l_3$ entanglement.
The link $l_1$ entanglement decoheres in 3 timesteps; the link $l_2$ entanglement decoheres in 2 timesteps; and the link $l_3$ entanglement decoheres after just 1 timestep.
Suppose for simplicity that request arrivals are Bernoulli with parameters $\lambda_{r_1},\lambda_{r_2},\lambda_{r_3}$ respectively and occur just before the first of these three-time slots. 

Notice in this setup, the $r_1$ and $r_2$ request queues are given the opportunity to schedule the LLEs in $l_1$ and $l_2$ before $r_3$. If we follow a myopic policy, like MaxWeight, then we will schedule these requests so long as the queues for $r_1$ or $r_2$ are non-empty. However, if we schedule queues $r_1$ or $r_2$, then we cannot serve requests in queue $r_3$. In other words, $r_1$ and $r_2$ have priority over $r_3$. Because we can only serve $r_3$ requests when there are no arrivals for $r_1$ and $r_2$, this induces the following necessary condition for stability under MaxWeight for the switch just described:
\[
\lambda_{r_3} < (1-\lambda_{r_1}) (1-\lambda_{r_2})\,.
\]
However, of course, we can plan ahead slightly, and in the last of the three timeslots, we can choose which requests to serve. This results in the following sufficient condition for stability :
\[
\lambda_{r_1} + \lambda_{r_3}< 1 ,\qquad \lambda_{r_2} + \lambda_{r_3} <1, \qquad \lambda_{r_3} <1 .
\]
Note that the three constraints above correspond, respectively, to the allocation of the three LLEs generated on links $l_1$, $l_2$, and $l_3$. Here, we can stabilize $\lambda_{r}=0.4$ $\forall r$, whereas MaxWeight is unstable.
From the inequalities above, we see that MaxWeight's stability region is strictly smaller than the capacity region for this quantum switch. 
This proof can be extended to settings to probabilistic decoherences. We state this in following theorem:
\begin{restatable}{thm}{MWInstable}\label{thrm:MWInstable}
    The MaxWeight policy is not optimal for throughput when LLEs have probabilistic decoherences.
\end{restatable}
This result is proven in Appendix~\ref{append:instability_maxweight}.
A simulation demonstrating the MaxWeight instability and stability of our alternative policy is given in Figure \ref{Fig:lyapunov_drift}.
So, MaxWeight is not throughput optimal. We will now demonstrate that this is because MaxWeight does not plan ahead. The logic behind MaxWeight is still valuable in designing throughput-optimal policies; however, we urge caution when directly implementing MaxWeight in quantum switches.

\subsection{Time-Scale Separation}
\label{sec:timescale}

In this section, we discuss timescale separation in a quantum switch. We also refer the reader to \cite{zubeldia2022matching} for an analysis of MaxWeight in a Quantum switch with $\textsc{Y}$ and $\textsc{W}$ matching topologies. Our discussion here helps us gain intuition that leads to an optimal policy. The formal demonstration of the observations made here is proven across Theorems \ref{thm:fluid_convergence}, \ref{thm:fluid_stability} and  \ref{thm:stochastic_stability}.

\begin{figure}
    \centering    
\subfigure[%
(MaxWeight instability): We set $\lambda_{r_1}=\lambda_{r_2}=0.005$, $\lambda_{r_3}=0.004$, $\lambda_{l_1}=\lambda_{l_2}=0.02$, $\lambda_{l_3}=0.01$, $d_{l_1}=d_{l_2}=0.00001$, $d_{l_3}=0.99999$, $\gamma_{r_i}=1$ for all $i\in\sbrac{3}$ and $c=200$. The optimal policy is computed through value iteration.%
        \label{Fig:lyapunov_drift}]{
    \centering
    \includegraphics[width=10cm]{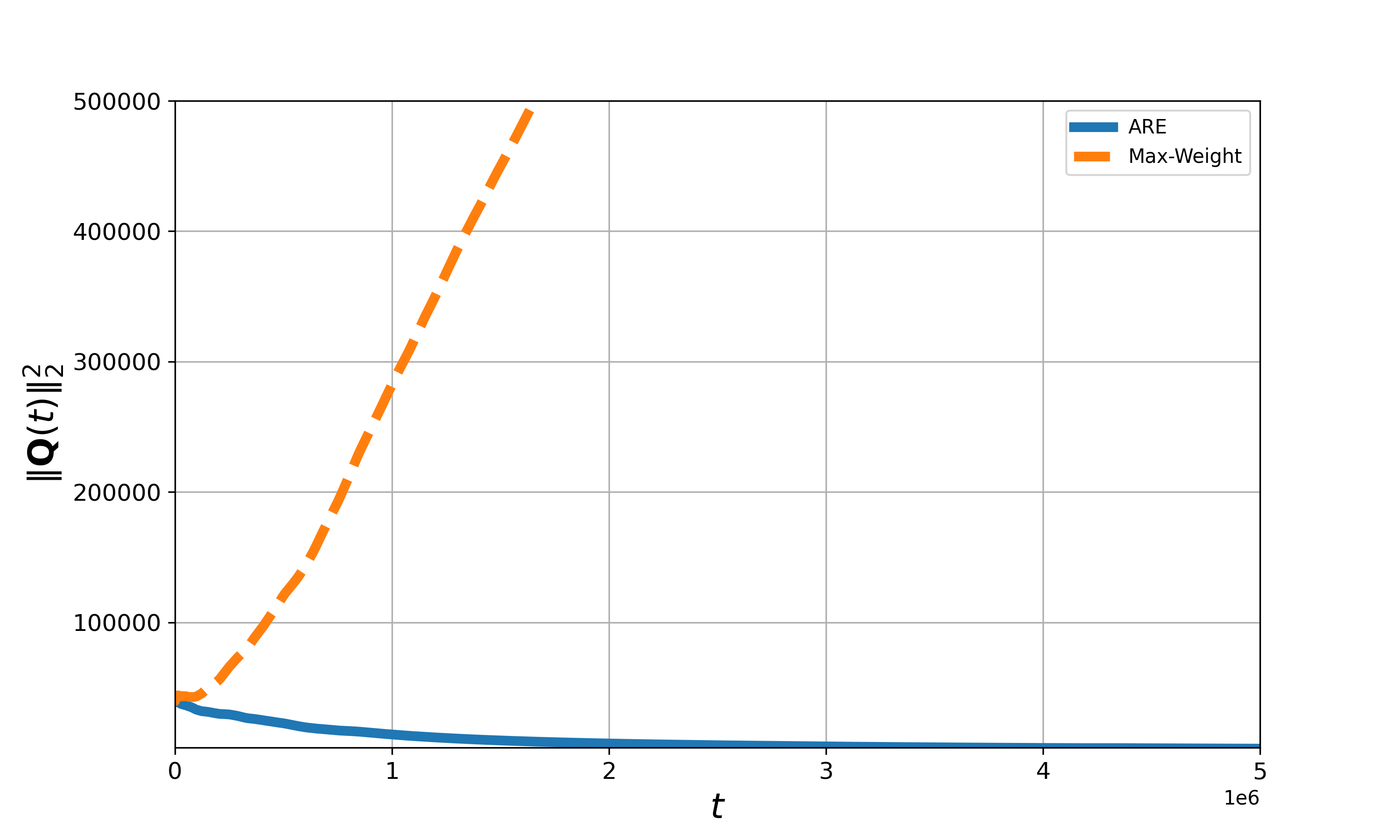}
        
        }
    \hfill
\subfigure[
(Time-scale separation): We set $\lambda_{r_1}=\lambda_{r_2}=0.05$, $\lambda_{r_3}=0.04$, $\lambda_{l_1}=\lambda_{l_2}=0.2$, $\lambda_{l_3}=0.1$, $d_{l_1}=d_{l_2}=0.00001$, $d_{l_3}=0.99999$, $\gamma_{r_i}=1$ for all $i\in\sbrac{3}$ and $c=200$. 
The ARE policy is used for LLE scheduling.%
\label{Fig:fl-qs}]{

    \centering
    \includegraphics[width=10cm]{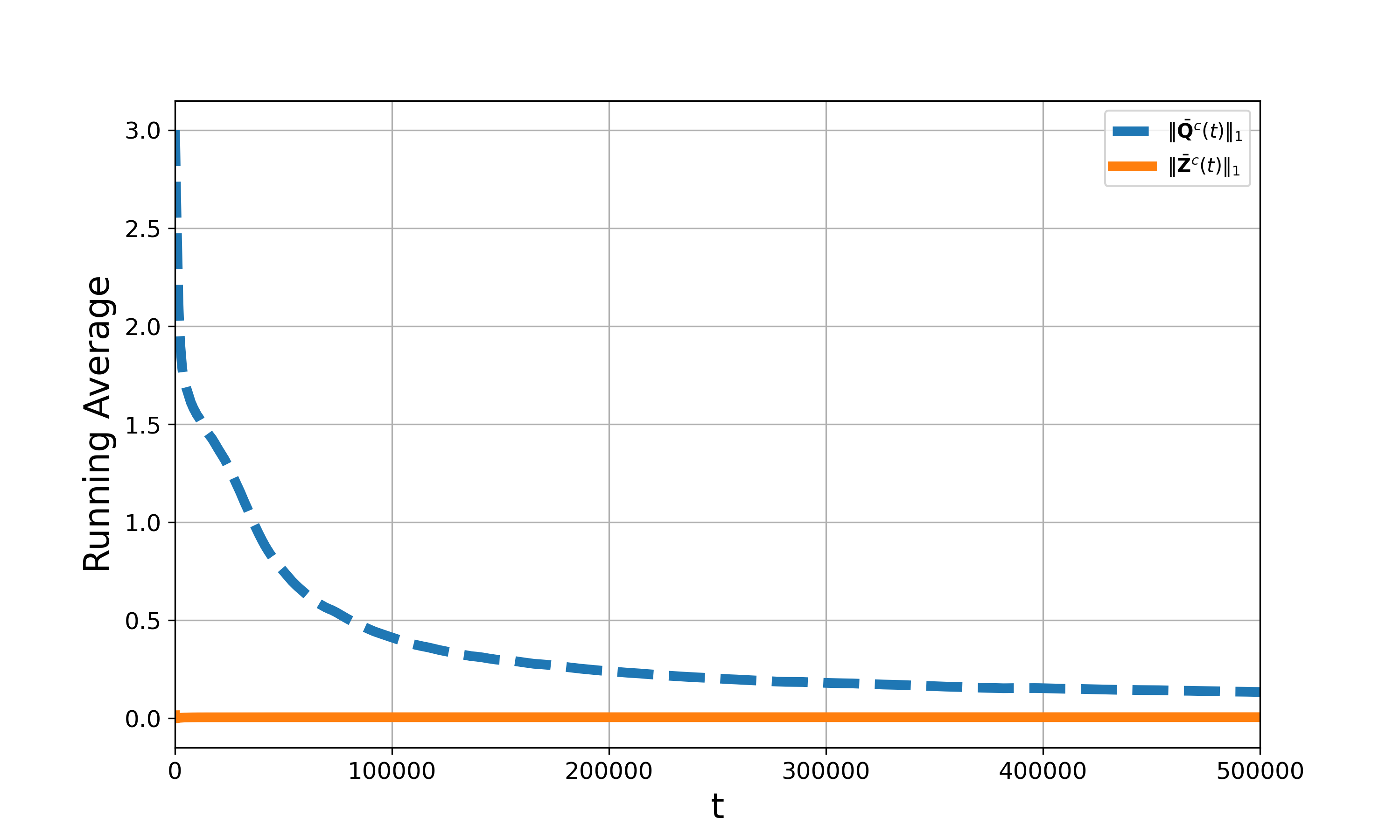}

}
\caption{
For simulations, we simulated the counter example given in Section~\ref{subsec:counter_example} with probabilistic decoherence of LLEs. In the counter example, we considered three types of requests ($\mathcal{R}=\cbrac{r_1,r_2,r_3}$) with three links ($\mathcal{L}=\cbrac{l_1,l_2,l_3}$). Moreover, we have considered $\mathcal{L}_{r_1}=\cbrac{l_1}$, $\mathcal{L}_{r_2}=\cbrac{l_2}$ and $\mathcal{L}_{r_3}=\cbrac{l_1,l_2,l_3}$. We set $B=1$.   
}
\label{Fig:comp}

\end{figure}

In Figure \ref{Fig:lyapunov_drift}, we plot the performance of MaxWeight against an alternative policy, ARE, which we will describe shortly. MaxWeight has optimal drift for the quadratic Lyapunov function in a classical switch. However, this is not the case for a quantum switch.
To evaluate the switch's performance, we must understand the time-scale separation that occurs under congestion. 

Consider the quantum switch where there are a large number of requests, i.e., $\sum_r Q_r(0) = c$ for $c\gg 1$. In this state, the number of the LLEs will be far smaller $\bm Z(t) = O(1)$.  Firstly, because of the physical limitations of the switch having bounded storage for LLEs. Second, because LLEs decohere, leading to much faster convergence in queue length. Timescale separation is demonstrated in Figure \ref{Fig:fl-qs}.
To make any relative change in the request process, we require $c$ transitions of the LLE process. Since the LLE process is already close to equilibrium being of order $O(1)$, it quickly converges to its stationary behavior. Thus, the resulting scheduling dynamics placed on the request queues are ultimately determined by the stationary behavior of the LLE process. (This further motivates the capacity region characterization that we already proved in Theorem \ref{thrm:stabilityRegion}.)





\subsection{Addressing the Sub-Optimality of MaxWeight}
In contrast, in classical switches, two-time scale separation affects the Lyapunov drift of the MaxWeight in such a way that it is no longer optimal. The original rationale of the MaxWeight policy is to achieve the maximum negative Lyapunov drift. We can informally explain this as follows:
The differential equation below approximates the request queues given in Section~\ref{sec:system_model},
\begin{equation*}
    \frac{d\bar{Q}_r(t)}{dt}=\lambda_r -  \gamma_r \mathbb E_{\bm z \sim \mu(t) } [n_r(t)], \qquad  r\in \mathcal{R}. \footnote{We more formally define $\bar Q$ and associated fluid model terms in Section \ref{sec:main_results}.}
\end{equation*}
Above $n_r(t)$ is the number of requests scheduled at time $t$. The number of requests in turn depends on the number of LLEs, $\bm z$, which has distribution $\mu(t)$ at stationarity. If
we take the function $L(\bar{\bm{Q}}(t)) = \sum_{r \in \mathcal{R}} \bar{Q}_r^2(t)/2$, which would be the Lyapunov function typically associated with MaxWeight. Then, differentiation with the chain rule gives
\[
 \frac{dL(\bar{\bm{Q}}(t))}{dt}= \sum_{r \in \mathcal R} \bar Q_r(t)\lambda_r -    \mathbb E_{\bm z \sim \mu(t) } \Big[ \sum_{r \in \mathcal R}\bar Q_r(t) \gamma_r  n_r(t)\Big]\, .
\]
The MaxWeight policy should maximize the negative drift of the Lyapunov function. However, the MaxWeight policy does not achieve the maximum negative drift for the quantum switch from the above equation. Specifically, the maximum negative drift solves the optimization  
\begin{equation}\label{eq:avMDP}
    \max_{\pi \in \mathcal A }\;\; \mathbb E_{\bm z \sim \mu^\pi } \Big[ \sum_{r \in \mathcal R}\bar Q_r(t) \gamma_r  n^\pi_r\Big] \, .
\end{equation}
In a classical switch or a wireless network, the above optimization is a combinatorial optimization problem, such as a bipartite matching problem. However, we see above that we must jointly optimize over schedules and their induced stationary distributions. In particular, the critical observation is that the above optimization is MDP. We discuss this point in more detail in the next subsection.

We note that the above argument is somewhat heuristic. To make this rigorous, we must prove the above time-scale separation and fluid limit are correct. We complete this in Theorem \ref{thm:fluid_convergence}. We also verify that the solution of the MDP gives the optimal drift, which we define below. Then, we investigate how this impacts stochastic policies that implement schedulers solving this MDP. This result is given in Theorem \ref{thm:fluid_stability}. \blue{ }

\subsection{Optimal Scheduling is an Average Reward MDP}
As introduced in the previous section, the optimal scheduling policy must solve an average reward MDP. 
For illustration, see Figure~\ref{Fig:lyapunov_drift}, which shows that the drift of the policy obtained as a solution of the average reward MDP is negative compared to that of the MaxWeight policy for the quantum switch, as depicted in Figure~\ref{Fig:sm_qs}. This negative drift under the average reward MDP solution reflects the policy's ability to more effectively balance short-term scheduling decisions with the long-term availability of resources.

We now briefly discuss average reward MDPs. For a detailed treatment, we refer the reader to Chapter 7 of Puterman \cite{puterman2014markov} or Chapter 5 of Bersekas \cite{bertsekas2011dynamic}.
The single-step reward associated with implementing schedule $ \bm n $, state $ \bm z $ and request queue length vector $\bm  q$ is
\begin{equation}
\label{eqn:reward}
 u(\bm{z}, \bm{n}) = \sum_{r \in \mathcal{R}} n_r \gamma_r q_r.   
\end{equation}
This reward represents the weighted average number of requests fulfilled in the current time slot. Notice in terms of the MDP framework, the system's state is $\bm z$, the action chosen is $\bm n$, and $\bm q$ is a parameter of our objective function. 
We assume that the states $\bm z$ evolve as described in Section \ref{sec:Qdynamics}. However, we take the queue size parameter $\bm q$ as a fixed constant, so our policy is myopic in that it does not account for the impact of decisions on the evolution of request queues.

We consider policies that are a function of the current link state $\bm z$.
The single-step reward associated with a policy $ \pi $ and state $ \bm{z} $ is given by:
$
u(\bm{z}, \pi(\bm{z})) = \sum_{(\bm \sigma(\bm n),\bm{n}) \in \mathcal{S}} \pi(\bm{n} \mid \bm{z}) \, u(\bm{z}, \bm{n}),
$
where $ \pi(\bm{n} \mid \bm{z}) $ represents the probability of choosing action $ \bm{n} $ given the state $ \bm{z} $ and request queue length under policy $ \pi $. 
Notice that under any policy, $\pi$ states evolve as a positive recurrent discrete-time Markov chain with a single irreducible communicating class. Thus, our average reward MDP is \emph{unichain}, which in turn implies the existence of stationary solutions to the average reward MDP \cite[Proposition 5.2.4]{bertsekas2011dynamic}.

The average reward associated with stationary policy $ \pi $ is denoted by $ R^{\pi} $ and is defined as:
\begin{equation*}
R^{\pi} = \lim_{T \to \infty} \frac{\mathbb{E} \left[\sum_{s=0}^{T-1} u(\bm{Z}(s), \pi(\bm{Z}(s))\right]}{T} = \mathbb E_{z\sim \mu^\pi} \Big[  \sum_{r \in \mathcal R} q_r  \gamma_r n_r^{\pi}(\infty) \Big] \, .
\end{equation*}
     Above, we note that the Markov chain $\bm Z(s)$ is ergodic, and we let $\mu^\pi$ denote the stationary distribution of $\pi$, and we let the random variable $n^\pi_r$ denote the stationary number of requests of type $r$ processed under $\pi$.\footnote{We assume that the policy $\pi$ will continue to process requests irrespective of whether requests are queued.} Rewriting the above expression in terms of these stationary quantities, the optimal reward and the optimal policy are denoted, respectively, by
\begin{equation} \label{eq:AREMDP}
R^{\star}(\bm q) = \max_{\pi}\,\, \mathbb E_{z\sim \mu^\pi} \Big[  \sum_{r \in \mathcal R} q_r  \gamma_r n_r^{\pi}(\infty) \Big]\, , \quad \pi^\star(\bm q) \in \argmax_{\pi}\,\, \mathbb E_{z\sim \mu^\pi} \Big[  \sum_{r \in \mathcal R} q_r  \gamma_r n_r^{\pi}(\infty) \Big]\, .    
\end{equation}

For each state $\bm{z} \in \mathcal Z$, the Bellman equation for the optimal reward $R^\star$ is given by:
\begin{equation}
\label{eqn:bellman}
R^\star(\bm q) + V^{\star}(\bm{z},\bm q) = \max_{(\bm \sigma(\bm n),\bm{n}) \in \mathcal{S}} \left\{ \sum_{r \in \mathcal R} q_r  \gamma_r n_r + \sum_{\bm{z}' \in \mathcal{Z}} P(\bm{z}' \mid \bm{z}, \bm{n}) V^{\star}(\bm{z}',\bm q) \right\},
\end{equation}
where $V^{\star}(\bm{z},\bm q)$ is the relative value function, representing the relative value of state $\bm{z}$ compared to the optimal-term average reward $R^\star(\bm q)$,  $P(\bm{z}' \mid \bm{z}, \bm{n})$ is the transition probability from the current state $\bm{z}$ to the next state $\bm{z}'$, given action $\bm{n}$.
The optimal policy maximizes~\eqref{eqn:bellman}, which by design balances the allocation of LLEs to satisfy current requests while also considering the impact on future system states. 
The optimal policy can be calculated through value iteration or policy iteration.  If the LLE process is not known, parameters can be learned directly or via tabular reinforcement learning algorithms, and these estimations can then be applied to calculated optimal scheduling decisions.

\subsection{The Average Reward Entanglement Scheduler}

This paper considers the optimal scheduling policy for allocating LLEs in the quantum switch, which is found by solving an average reward MDP that optimizes the utilization of LLEs to fulfill incoming requests. 

We call this policy the Average Reward Entanglement (ARE) scheduler. It consists of a single parameter $\tau \in\mathbb N$ and the sequential implementation of the average reward MDP described above. 
The ARE policy is defined as follows:
\begin{quote}
        \textbf{ARE}: At each time $t$ divisible by $\tau$, the scheduler observes the state of the right-side process, denoted by $ \bm{Q}(t)$. The policy then implements the scheduler $\pi^\star(\bm Q(t))$ defined by \eqref{eq:AREMDP} for time steps $t,t+1,...t+\tau-1$\,.
\end{quote}

The above policy is natural for a quantum switch, where generated LLEs decohere much faster than connection-level requests.
Another vital point on the ARE policy is that it does not need to know the average arrival rate $\bm \lambda$. 
The ARE policy makes decisions where current queue sizes are simply a parameter used to determine how many the LLEs are created, scheduled and decohere regardless of the state evolution of the request queues. (I.e., if requests are requested to serve, ARE will serve them. However, even if a request queue is empty, the policy will continue to schedule jobs regardless of the availability of requests.) Since we only require current queue sizes, the ARE policy behaves similarly to the Max-Weight policy, which doesn’t require explicit knowledge of $\bm \lambda$ to achieve optimal throughput.

While traffic patterns do not need to be learned, the policy requires the solution of an MDP. 
MDPs are well-known to suffer from the curse of dimensionality, in our case as the buffer size grows. 
In practice, reinforcement learning methods and function approximation would be implemented off-line to find the switch's optimal dynamics prior to implementation. These directions would be an important area for future investigation.

\section{Main Results}
\label{sec:main_results}
This section presents and discusses the main results of the ARE policy. For simplicity in notation, let $\pi^{\star}(\tau(c))$ denote the ARE policy for a fixed $c \in \Nats$.

To demonstrate that the ARE policy is asymptotically throughput optimal,
 we consider a sequence of ARE schemes $\brac{\pi^{\star}(\tau(c))}_{c \in \Nats}$, where for every $c \in \Nats$, the scheme $\pi^{\star}(\tau(c))$ is obtained by~\eqref{eq:AREMDP}. 
We first establish the fluid limit of the quantum switch under the ARE policy (Theorem~\ref{thm:fluid_convergence}). 
Furthermore, we show that the fluid model corresponding to the ARE policy is throughput optimal (Theorem~\ref{thm:fluid_stability}). Finally, we establish the asymptotic throughput optimality of the ARE policy (Theorem~\ref{thm:stochastic_stability}).


\subsection{Fluid Limit}
To establish the fluid limit, we define the scaled processes $\bar{\bm X}^c=\brac{\bar{\bm X}^c(t): t\in \Z_+}$ where $\bar{\bm X}^c(t)=\brac{\bar{\bm Q}^c(t),\bar{\bm Z}^c(t)}$  is such that
$
\bar{\bm Q}^c(t)=\frac{\mf Q(ct)}{c}, \  \bar{\bm Z}^c(t)=\frac{\mf Z(ct)}{c}, \forall t\in \Z_+, \ c\in \Nats.
$
Here, we accelerate time by the scaling factor $c$ and scale space by this same factor.
An important facet is the two-time scale separation between the processes $\bar{\bm Z}^c$  and $\bar{\bm Q}^c$ as discussed in Section \ref{sec:timescale}. 
In the limit as $c \to \infty$, the $\bar{\bm Z}^c$  process is zero, yet still has a meaningful impact on the evolution of the process $\bar{\bm Q}^c$.
Further in the sequence above, we allow the arrival rate $\bm \lambda^c$ to depend on $c$.

We now focus on the fluid model for the ARE policy defined as follows.
\begin{defn}(ARE Fluid model)
\label{defn:fluid_limit}
Given an initial condition $\bar{\bm Q}(0)\in \R_+^{R}$ such that $\norm{\bar{\bm Q}(0)}_1>0$, a Lispchitz continuous function $\bar{\bm Q}:[0,\infty) \rightarrow \R_+^{R}$ is said to be a solution to the fluid model if it satisfies following equations for all $t,s\geq0$ and $r\in\mathcal{R}$ 
\begin{align}
\label{eqn:fl_1}
\bar{Q}_r(t)&=\bar{Q}_0(t)+ \bar{A}_r(t)- \bar{D}_r(t),\\
\bar{A}_r(t)&=\lambda_r t,\label{eqn:fl_2}\\
\int_{s}^t \sum_{r \in \mathcal R}\bar{Q}_r(u) d \bar{D}_r(u) &\geq \max_{\bar{\bm \pi}\in \mathcal A} \int_s^t \sum_{r \in \mathcal R}\bar{Q}_r(u) \E_{z\sim\mu^{\bar{\bm \pi}}}[\gamma_r \bar{n}_r(\infty)]du
 \label{eqn:fl_3},
\end{align}
where 
$\mu^{\bar{\bm \pi}}$ denotes the steady-state distribution of the left-side process $\bm{Z}$ under policy $\bar{\bm \pi}$ and $\bar D_r(t)$ is increasing and Lipschitz continuous.
\end{defn}
The following result is somewhat technical. It establishes that limits exist for sequences $\brac{\bar{\bm Q}^c}_{c}$, and when they exist, it proves that they must satisfy the fluid equations above. Thus, this proves that our fluid limit model equations represent the asymptotic behavior of our queueing process. 
\begin{thm}[Fluid Limit]
\label{thm:fluid_convergence}
For $\tau(c)=(\log(c))^2$ and $\bar{\bm Q}(0) \in \R_+^{|\mathcal{R}|}$ with $ \norm{\bar{\bm Q}^c(0)}_1 \leq 1 $ and $ \bar{\bm Q}^c(0){\to} \bar{\bm Q}(0) $ a.s. and in $\ell_1$ and $\bm \lambda^c \rightarrow \bm \lambda$  as $c \rightarrow \infty$. Then the sequence of stochastic processes $\brac{\bar{\bm{Q}}^c}_{c \in \mathbb{N}}$ under the ARE policy is tight with respect to the topology of uniform convergence on compact time intervals. Additionally, every weakly convergent subsequence of $\brac{\bar{\bm{Q}}^c}_{c \in \mathbb{N}}$ converges to a Lipschitz continuous process $\bar{\bm Q}$. This limiting process satisfies the fluid model equations~\eqref{eqn:fl_1}-\eqref{eqn:fl_3}.
\end{thm}
Hence, the process $\bar{\bm Q}$, defined in Theorem~\ref{thm:fluid_convergence}, characterizes the dynamics of the quantum switch under the ARE policy in the limit as $c \to \infty$. 

It is important to note that the scheduling schemes for quantum or classical switches can respond relatively quickly to changes in request queue lengths. By setting $\tau(c) = (\log(c))^2$, the system allows sufficient time for the link process to reach its steady state between changes in request queue lengths. Here, we use this choice since the mixing time of the LLE process is in the order of $\log(c)$. 
Setting $\tau$ to grow as $(\log (c))^2$ ensures that request queue updates are not too frequently affecting link process while maintaining efficiency.

In our following result, we show that for the fluid limit solutions of every weakly convergent subsequence $\brac{\bar{\bm{Q}}^c}_{c \in \Nats}$, the schedule $(\bm \sigma(\bm n^\star), \bm n^\star)$ is selected by the policy $\bar{\pi}^\star$, which is the limit of the sequence of policies $\brac{\pi^\star(\tau(c))}_{c \in \Nats}$, and is obtained from~\eqref{eq:AREMDP}.

\subsection{Fluid Limit Stability.}
Our fluid model is considered to be stable if there exists a $T>0$ such that for every $\bar{\bm Q}$ satisfying equations~\eqref{eqn:fl_1}-\eqref{eqn:fl_3} with $\norm{\bar{\bm Q}(0)}_1\neq0$, we have
\begin{equation}
\bar{Q}_r(t)=0, \ r\in \mathcal{R}, \forall \ t\geq T.
\end{equation}
We say a fluid limit is throughput optimal if it is stable for all $\bm \lambda \in \mathcal C^\circ$. 
The following result proves that the fluid model corresponding to the ARE policy $\bar{\pi}^\star$ is optimal throughput.
\begin{thm}
\label{thm:fluid_stability}
The ARE fluid model (\ref{eqn:fl_1}-\ref{eqn:fl_3}) is throughput optimal.
\end{thm}

\subsection{Stochastic Stability.}
Using fluid limit stability, we see the process $\bar{\bm X}^c$ is asymptotically throughput optimal.

\begin{thm}
\label{thm:stochastic_stability}
The ARE policies are asymptotically throughput optimal.
\end{thm}
\noindent So by specifying how mixed the LLE queues are, we can specify any level of stability up to the full capacity region. 
The proof of Theorem \ref{thm:stochastic_stability} in given in Section \ref{append:stochastic_stability}
 in the Appendix.


\section{Proofs of Main Results}
\label{sec:proofs}
This section presents the primary proofs of the results discussed in Section~\ref{sec:main_results}.
\subsection{Fluid Limit: Proof of Theorem~\ref{thm:fluid_convergence}}
The proof of Theorem~\ref{thm:fluid_convergence} is structured as follows.
We show that the sequence of stochastic processes $\brac{\bar{\bm{Q}}^c}_{c \in \Nats}$ has a subsequence with a limit. This argument is standard -- albeit somewhat technical -- and employs the Arzelà-Ascoli Theorem~\cite{billingsley2013convergence}. 
With the existence of limits confirmed, we explore their properties.
In particular,
 in Proposition \ref{prop:timescale},
 we examine the timescale separation of $\bar{\bm{Q}}^c$ and $\bar{\bm{Z}}^c$ over the time interval $ct$ by partitioning time into time intervals of length $\tau(c)$. 
  Within each length $\tau(c)$ interval, the schedules $(\bm{\sigma}(\bm{n}), \bm{n})$ selected by policy $\pi^\star(\tau(c))$ are determined based on the request queue lengths at the start of the segment. We can then apply this in Proposition \ref{prop:opt} to give fluid equation \eqref{eqn:fl_3}.

\subsubsection{Tightness}
\label{subsec:tightness}
We first define the family of coupled processes $\brac{\bar{\bm{Q}}^c, \bar{\bm{A}}^c, \bar{\bm{D}}^c }$, where $\bar{\bm{Q}}^c(t)$, $\bar{\bm{A}}^c(t)$, and $\bar{\bm{D}}^c(t)$ are the scaled versions of the request queue process, the cumulative arrivals and departures, respectively. These processes are constructed on the same probability space and remain the same for different values of $c$.
For each $r \in \mathcal{R}$ and $l\in \mathcal{L}$, we define the scaled processes for arrivals, departures and LLE queues as:
\begin{equation}
\bar{A}_r^c(t)=\frac{ A_r(ct)}{c}, \quad\bar{D}_r^c(t)=\frac{ D_r(ct)}{c},\quad \bar{Z}_l^c(t)=\frac{Z_l(ct)}{c},\quad \ c\in \Nats.
\end{equation}

The proof of tightness is contained in the following proposition.

\begin{restatable}{prop}{PropositionTightness}\label{prop:Tightness}
       The sequence of stochastic processes $\brac{\bm \lambda^c,\bar{\bm{Q}}^c,\bar{\bm{A}}^c,\bar{\bm{D}}^c, \bar{\bm Z}^c}_{c \in \mathbb{N}}$ under the ARE policy is tight with respect to the topology of uniform convergence on compact time intervals.  
\end{restatable}

\noindent The proof of Proposition \ref{prop:Tightness} is given in Section~\ref{app:tightness} in the appendix.
Tightness implies relative compactness \cite[Prohorov's Theorem]{billingsley2013convergence}: that for every $\brac{\bm \lambda^c,\bar{\bm{Q}}^c,\bar{\bm{A}}^c,\bar{\bm{D}}^c, \bar{\bm Z}^c}_{c \in \mathbb{N}}$ there is a weakly convergent sub-sequence.
This verifies the tightness statement in Theorem \ref{thm:fluid_convergence} and also fluid limit equation \eqref{eqn:fl_1}.
Applying the Skorohod Representation Theorem \cite{billingsley2013convergence} we may also assume that this sub-sequence convergence holds almost surely. 
Throughout the rest of the proof we assume that $\brac{\bm \lambda^c,\bar{\bm{Q}}^c,\bar{\bm{A}}^c,\bar{\bm{D}}^c, \bar{\bm Z}^c}_c$ is any such convergence sub-sequence and its limit point is $\brac{\bm \lambda,\bar{\bm{Q}},\bar{\bm{A}},\bar{\bm{D}}, \bar{\bm Z}}\,$. By applying the Functional Strong Law of Large Numbers, $\bar{A}_r^c(t)$ converges uniformly on compacts to $ \lambda_r t$  as $c \to \infty$ this verifies \eqref{eqn:fl_2}. It remains to prove \eqref{eqn:fl_3}.

\subsubsection{Timescale Separation}
To prove \eqref{eqn:fl_3}, we first need to understand the timescale separation behavior of these switches on the fluid scale. The main finding of this section is summarized in the following proposition.

\begin{restatable}{prop}{TimeScale}\label{prop:timescale}
 For any interval $[u,t]$ for which $\bar Q_r(s)>0$ for all $s \in [u,t]$ the following holds
\begin{equation}
\bar{D}_r(t) - \bar{D}_r(u)  = \lim_{c\rightarrow \infty} \frac{1}{c} \sum_{s=cu}^{ct} \E_{z\sim \mu^{\pi^\star}(s)} \sbrac{\gamma_r {n}^\star_r}\,,
\label{eq:Dlim}    
\end{equation}
where above $\mu^{\pi^\star}(s)$ is the stationary distribution of the LLEs under the ARE policy at time $s$ and $n_r^\star$ is the stationary number of type $r$ entanglement swaps under $\mu^{\pi^\star}(s)$.   
\end{restatable} 

\begin{proof}
To simplify notation in the proof, without loss of generality, we can shift time and assume that $u=0$.
Since $\bar{Q}_r(s)>0$ for $s\in [0,t]$ and $\bar{Q}^c_r(s)$ convergences uniformly to $\bar{Q}_r(s)>0$, there exists a value of $c'$ for which $\bar{Q}^c_r(s)$ is strictly bounded away from zero for all $c\geq c'$. Since queues are non-zero,
we can write $\bar{D}^c_r(t)$ as
\begin{align}
\bar{D}^c_r(t)&= \frac{1}{c}\sum_{s=1}^{ct}\hat{n}^\star_r\brac{\bm Z(s),\bar{\bm Q}^c(s/c)}\label{eqn:scales_Q}\, .
\end{align}
Also, as described above, $\bar{Q}^c_r(0)$ converges to $\bar{Q}_r(0)$.
Here, we explicitly emphasize the dependence of the number of successful requests, $\hat{\bm n}^\star$, on both the link process $\bm Z$ and the request queue process $\bm Q$. Note that the policy $\pi^\star(\tau(c))$ is obtained as a solution of average reward MDP, and these solutions do not change if we rescale the queue size vector by a factor $c$; that is, we have
$\hat{n}^\star_r\brac{\bm Z(s),{\bm Q}(s)}=\hat{n}^\star_r({\bm Z(s),\bar{\bm Q}^c(s/c)}).$

The departure term, equation~\eqref{eqn:scales_Q}, is now decomposed over equal intervals of length $\tau(c)$. This decomposition is beneficial as it leverages the time-scale separation between the link and request processes. More precisely, the departure term can be written as\footnote{Note that if $ct$ is not divisible by $\tau(c)$, there will be an error term corresponding to the small time interval. However, this error term is of the order $ O(1/c)$, which tends to zero as $ c \to \infty $.}:
\begin{align}
 &\bar{D}^c_r(t)\nonumber
\\ &= 
 \frac{1}{c} \sum_{i:i\tau(c) \leq ct} \Bigg[\sum_{s=i\tau(c)}^{(i+1)\tau(c)-1}\hat{n}^\star_r\brac{\bm Z(s),\bar{\bm Q}^c(i\tau(c)/c)} \nonumber\\ 
 &\quad \quad\quad\quad \quad\quad-\sum_{s=i\tau(c)}^{(i+1)\tau(c)-1} \E \sbrac{\hat{n}^\star_r\brac{\bm Z(s),\bar{\bm Q}^c(i\tau(c)/c)} \mid \bm X(i\tau(c))}\Bigg]\label{eqn:term1} \\
 &+ \frac{1}{c} \sum_{i:i\tau(c) \leq ct} \Bigg[\sum_{s=i\tau(c)}^{(i+1)\tau(c)-1} \Big(\E \sbrac{\hat{n}^\star_r\brac{\bm Z(s),\bar{\bm Q}^c(i\tau(c)/c)} \mid \bm X(i\tau(c))}\nonumber \\
 & \quad \quad\quad\quad- \E \sbrac{\hat{n}^\star_r\brac{\bm Z(s),\bar{\bm Q}^c(i\tau(c)/c)} \mid \bm Z(i\tau(c))\sim \mu^{\pi^\star(\bm Q(i\tau(c))} ,\bm Q(i\tau(c))}  \Big)\Bigg] \label{eqn:term2}  \\
 &+\frac{1}{c} \sum_{i:i\tau(c) \leq ct} \sum_{s=i\tau(c)}^{(i+1)\tau(c)-1} \E \sbrac{\hat{n}^\star_r\brac{\bm Z(s),\bar{\bm Q}^c(i\tau(c)/c)}\mid \bm Z(i\tau(c))\sim \mu^{\pi^\star(\bm Q(i\tau(c))} ,\bm Q(i\tau(c))}\label{eqn:term3}\\
 & + o(1) \nonumber.
\end{align}

The plan for the remainder of the proof is as follows: for term~\eqref{eqn:term1}, the Azuma-Hoeffding inequality will be used to show that it approaches zero as $c \to \infty$. 
Moreover, term~\eqref{eqn:term2} represents the difference between the expected departures when the process $\bm{Z}$ starts from its stationary distribution $\mu^{\pi^\star(\bm Q(i\tau(c))}$ under the policy $\pi^\star(\bm Q(i\tau(c))$ as defined by \eqref{eq:AREMDP} and when $\bm{Z}$ begins from an arbitrary distribution. Leveraging time-scale separation shows that term~\eqref{eqn:term2} vanishes as $c \to \infty$.
From this we will see that the convergence of $\bar{D}_r^c(t)$ term~\eqref{eqn:term3} is determined by the term ~\eqref{eqn:fl_3} as $c \to \infty$. So, we now analyze terms~\eqref{eqn:term1}-\eqref{eqn:term3} separately.

 First for term \eqref{eqn:term1}, we note that, for $s=0,...,\tau(c)-1$,
$$
\sum_{i:i\tau(c) \leq ct} \sbrac{\hat{n}^\star_r\brac{\bm Z(i\tau(c)+s),\bar{\bm Q}^c(i\tau(c)/c)}-\E \sbrac{\hat{n}^\star_r\brac{\bm Z(i\tau(c)+s),\bar{\bm Q}^c(i\tau(c)/c)} \mid \bm X(i\tau(c))}},
$$
forms a martingale difference sequence. Therefore, in Lemma~\ref{lem:azzuma_hoeffding}, we apply the Azuma-Hoeffding inequality to prove that the term of~\eqref{eqn:term1} converges to zero as $c\to \infty$.

Now, we turn our attention to term~\eqref{eqn:term2}. We can further decompose~\eqref{eqn:term2} as follows:
\begin{align}
&\left|\eqref{eqn:term2}\right|
\notag
\\
&\leq
\frac{1}{c} \sum_{i:i\tau(c) \leq ct}\Bigg| \sum_{s=i\tau(c)}^{i\tau(c)+\tau_{\mix}(c)-1} \Big(\E \sbrac{\hat{n}^\star_r(s) \mid \bm X(i\tau(c))} \nonumber \\
&\qquad \qquad \qquad \qquad \qquad \qquad- \E \sbrac{\hat{n}^\star_r(s) \mid \bm Z(i\tau(c))\sim \mu^{\pi^\star\bm Q(i\tau(c))}  ,\bm Q(i\tau(c))}  \Big)\Bigg| \label{eqn:inter_1}\\
&
+\frac{1}{c}\sum_{i:i\tau(c) \leq ct} \Bigg|\sum_{s=i\tau(c)+\tau_{\mix}(c)}^{(i+1)\tau(c)-1} \Big(\E \sbrac{\hat{n}^\star_r(s) \mid \bm X(i\tau(c))} \nonumber\\
&\qquad \qquad \qquad \qquad \qquad \qquad- \E \sbrac{\hat{n}^\star_r(s) \mid \bm Z(i\tau(c))\sim \mu^{\pi^\star\bm Q(i\tau(c))}  ,\bm Q(i\tau(c))}  \Big)\Bigg|\label{eqn:inter_2},
\end{align}
for $0\leq \tau_{\mix}(c) \leq \tau(c)$.  It is important to chose $\tau_{\mix}(c)$ so that the link process has (approximately) reached its steady state when the system reaches $\tau_{\mix}(c)$. Therefore, until $\tau_{\mix}(c)$, there is a need to sacrifice system performance. It will be shown that the payoff up to $\tau_{\mix}(c)$, corresponding to term~\eqref{eqn:inter_1}, is bounded. 
Using the triangle inequality and observing that the maximum number of requests served in a single time slot is bounded by 
$B$,~\eqref{eqn:inter_1} can be upper-bounded as follows:

\begin{align}
&\frac{1}{c} \sum_{i:i\tau(c) \leq ct}\Bigg| \sum_{s=i\tau(c)}^{i\tau(c)+\tau_{\mix}(c)-1} \Big(\E \sbrac{\hat{n}^\star_r(s) \mid \bm X(i\tau(c))} \nonumber\\
&\qquad \qquad \qquad \qquad \qquad \qquad- \E \sbrac{\hat{n}^\star_r(s) \mid \bm Z(i\tau(c))\sim \mu^{\pi^\star\bm Q(i\tau(c))}  ,\bm Q(i\tau(c))}  
\Big)\Bigg|
\notag
\\
&\leq  \frac{2 B \tau_{\mix}(c)t}{\tau(c)}.    \label{eqn:t_1_term_bound}
\end{align}
Now for \eqref{eqn:inter_2}, we use the time-scale separation to show it vanishes as $c\to\infty$. By leveraging the assumption of policy $\pi^\star(\tau(c))$ that during each time segment $\cbrac{i\tau(c),\dots,(i+1)\tau(c)}$, the schedules $(\bm{\sigma}(\bm{n}(j)), \bm{n}(j))$ for all $j\in\cbrac{i\tau(c),\dots,(i+1)\tau(c)-1}$ are determined based on the request queue lengths at the beginning of the segment that is $\bar{\bm Q}^c(i\tau(c))$, even though request queue lengths evolve during this time segment. Using this, we can rewrite and upper bound~\eqref{eqn:inter_2} as follows:
\begin{align}
\eqref{eqn:inter_2}&\leq \frac{1}{c}
\sum_{i:i\tau(c) \leq ct} 
\sum_{s=i\tau(c)+\tau_{\mix}(c)}^{(i+1)\tau(c)-1}
    \Bigg| 
        \mu'_{\bm Q(i\tau(c))} \sbrac{P^{\bm Q(i\tau(c))}}^s \hat{n}^\star_r\brac{\bm Z(s),\bar{\bm Q}^c\Big(\frac{i\tau(c)}{c}\Big)}\\
        & \qquad \qquad \qquad \qquad \qquad \qquad-\mu^{\pi^\star(\bm Q(i\tau(c))} \sbrac{P^{\bm Q(i\tau(c))}}^s \hat{n}^\star_r\brac{\bm Z(s),\bar{\bm Q}^c\Big(\frac{i\tau(c)}{c}\Big)} 
    \Bigg| \nonumber\\
& \leq\frac{1}{c}\sum_{i:i\tau(c) \leq ct} \sum_{s=i\tau(c)+\tau_{\mix}(c)}^{(i+1)\tau(c)-1} 2B \norm{\mu'_{\bm Q(i\tau(c))} \sbrac{P^{\bm Q(i\tau(c))}}^s -\mu^{\pi^\star(\bm Q(i\tau(c))} \sbrac{P^{\bm Q(i\tau(c))}}^s}_{\TV}\label{eqn:inter_3}\\
&\leq \frac{1}{c}\sum_{i:i\tau(c) \leq ct} \sum_{s=i\tau(c)+\tau_{\mix}(c)}^{(i+1)\tau(c)-1} 2B \norm{\mu'_{\bm Q(i\tau(c))}  -\mu^{\pi^\star(\bm Q(i\tau(c))} }_{\TV} \rho^s \label{eqn:inter_4}\\
&\leq \frac{4B\rho^{\tau_{\mix}(c)}t}{(1-\rho)\tau(c)}\label{eqn:inter_5}.
\end{align}
In the first inequality, $ \mu'_{\bm{Q}(i\tau(c))}$ represents the initial distribution of the $\bm{Z}$ process which may not be stationary.
Furthermore, \eqref{eqn:inter_3} is derived from: 
\begin{multline*}
 \left| \mu'_{\bm Q} \sbrac{P^{\bm Q}}^s \hat{n}^\star_r(i\tau(c))-\mu^{\pi^\star(\bm Q)}\sbrac{P^{\bm Q}}^s \hat{n}^\star_r(i\tau(c)) \right| 
 \leq B     \sum_{\bm z\in \mathcal{Z}} \left|\sum_{\bm z'\in \mathcal{Z}}\brac{\mu'_{\bm Q}(\bm z') -\mu^{\pi^\star(\bm Q)}}\sbrac{P^{\bm Q}}^s_{\bm z' \bm z}\right|
 \\= 2B     \frac{1}{2}\sum_{\bm z\in \mathcal{Z}} \left|\mu'_{\bm Q}(\bm z)\sbrac{P^{\bm Q}}^s_{ \bm z} -\mu^{\pi^\star(\bm Q)}(\bm z)\sbrac{P^{\bm Q}}^s_{ \bm z}\right|=2B \norm{\mu'_{\bm Q}\sbrac{P^{\bm Q}}^s-\mu^{\pi^\star(\bm Q)}\sbrac{P^{\bm Q}}^s}_{\TV},
\end{multline*}
(For brevity, we suppress the dependence of $\bm{Q}$ on $ i\tau(c)$). In~\eqref{eqn:inter_4}, we have used our mixing time lemma, Lemma~\ref{lem:lemrho}. Finally, in the last inequality~\eqref{eqn:inter_5}, we used the bound $\norm{\mu'_{\bm{Q}(i\tau(c))} - \mu^{\pi^\star(\bm Q(i\tau(c))}}_{\TV} \leq 2$.

Therefore, applying~\eqref{eqn:t_1_term_bound} and~\eqref{eqn:inter_5} to \eqref{eqn:term2}, we have 
\begin{equation*}
\left|\eqref{eqn:term2}\right|\leq \frac{2 B \tau_{\mix}(c)t}{\tau(c)}+ \frac{4B\rho^{\tau_{\mix}(c)}t}{(1-\rho)\tau(c)}.
\end{equation*}
Hence, by selecting $\tau_{\mix}(c) =\left|\frac{\alpha\log(c)}{\log(\rho)}\right|$ for $\alpha>0$ and choosing $\tau(c) = (\log(c))^2$, the upper bound in the above expression approaches $0$ as $c \to \infty$. This proves that~\eqref{eqn:term2} converges to $0$ as $c\to \infty$.

Finally we turn to the limit of~\eqref{eqn:term3}. Since we are staring $Z(i\tau(c))\sim \mu^{\pi^\star(\bm Q(i\tau(c))}$, we can write
\begin{multline*}
\frac{1}{c} \sum_{i:i\tau(c) \leq ct} \sum_{s=i\tau(c)}^{(i+1)\tau(c)-1} \E \sbrac{\hat{n}^\star_r\brac{\bm Z(s),\bar{\bm Q}^c(i\tau(c)/c)}\mid \bm Z(i\tau(c))\sim \mu^{\pi^\star(\bm Q(i\tau(c))} ,\bm Q(i\tau(c))}=\\\frac{1}{c} \sum_{i:i\tau(c) \leq ct} \sum_{s=i\tau(c)}^{(i+1)\tau(c)-1} \E_{z\sim \mu^{\pi^\star(\bm Q(i\tau(c))}} \sbrac{\gamma_r {n}^\star_r}.
\end{multline*}

So we have established that Term~\eqref{eqn:term1} tends to zero as $c \to \infty$ by applying the Azuma-Hoeffding inequality, and leveraging time-scale separation, term~\eqref{eqn:term2} also vanishes as $c \to \infty$. So term~\eqref{eqn:term3} is the only term remaining. 
Thus we see that
\begin{equation}
\bar{D}_r(t) = \lim_{c\rightarrow \infty} \frac{1}{c} \sum_{i:i\tau(c) \leq ct} \sum_{s=i\tau(c)}^{(i+1)\tau(c)-1} \E_{z\sim \mu^{\pi^\star(\bm Q(i\tau(c))}} \sbrac{\gamma_r {n}^\star_r}\,.
\label{eq:Dlim2}    
\end{equation}
Thus, we see that \eqref{eq:Dlim} holds as required.  
\end{proof}

Here, we see that due to timescale separation, the departure process on the fluid scale is ultimately determined by the stationary distribution of LLEs. This observation is critical both in characterizing the non-standard capacity region of these system and for finding throughput optimal policies.


We now see that the stationary distribution of the ARE policy determines the departure process's limit.  In the next section, we can use this to establish the departure process's optimality.


\subsubsection{Optimality of Limit Points}

So far, we have established that the limit of any convergent subsequence of $\brac{\bar{\bm Q}^c}_{c\in \Nats}$ satisfies fluid model equations~\eqref{eqn:fl_1}-\eqref{eqn:fl_2} and we have seen that the departure process is determined by a timescale separation. We now need to verify \eqref{eqn:fl_3} with the following proposition. 
\begin{restatable}{prop}{OptProp}\label{prop:opt} For any agnostic policy $\bar{\pi}\in \mathcal A$ the following inequality holds
  \begin{equation}
\label{eqn:mdp_optimal}
    \int_s^t \sum_{r \in \mathcal R} \bar{ Q}_r(u) 
     d\bar{D}_r(u)\geq \int_s^t \sum_{r \in \mathcal R}\bar{ Q}_r(u)  \E_{z \sim \mu^{\bar{\pi}}}[\gamma_r{{ n}_r}]du\, .
     \tag{\ref{eqn:fl_3}}
\end{equation}  
\end{restatable}
\noindent Proposition \ref{prop:opt} is a consequence of Proposition \ref{prop:timescale}. Its proof is given in Section \ref{sec:proofopt} of the appendix.


We can now conclude the proof of Theorem \ref{thm:fluid_convergence}. We have now shown, as required, the tightness of the sequence $\brac{\bm \lambda^c,\bar{\bm{Q}}^c,\bar{\bm{A}}^c,\bar{\bm{D}}^c, \bar{\bm Z}^c}_{c \in \mathbb{N}}$ and that any limit point of this sequence satisfies the fluid model equations \eqref{eqn:fl_1}, \eqref{eqn:fl_2} and \eqref{eqn:fl_3}. This completes the proof  of Theorem \ref{thm:fluid_convergence}.\hfill $\blacksquare$

\subsection{Fluid Stability: Proof of Theorem~\ref{thm:fluid_stability}}
To prove the fluid stability, we show that there exists a $T>0$ such that for every fluid solution $\brac{\bar{\bm A},\bar{\bm Q},\bar{\bm D}}$ satisfying equations~\eqref{eqn:fl_1}-\eqref{eqn:fl_3} with $\norm{\bar{\bm Q}(0)}_1 = 1$, we have
$
\bar{Q}_r(t)=0, \ r\in \mathcal{R}, \forall \ t\geq T.
$
Consider a Lyapunov function $L(\bar{\bm Q}(t)) = \frac{1}{2}\sum_{r \in \mathcal R}\bar{Q}_r^2(t).$
Using fluid model equations~\eqref{eqn:fl_1}-\eqref{eqn:fl_3}, 
the derivative of $L(\bar{\bm Q}(t))$ is:
\begin{align}
    \frac{dL(\bar{\bm Q}(t))}{dt}&= \sum_{r\in \mathcal{R}}\bar{Q}_r(t) \lambda_r -\sum_{r\in \mathcal{R}} \bar{Q}_r(t) \bar{D}'_r(t)
    \label{eqn:lyapunov_drift}.
\end{align}
Now from Theorem~\ref{thrm:stabilityRegion}, we know that for any $\bm \lambda \in \mathcal{C}^\circ$ there exists an agnostic policy $\pi'$ such that for some $\epsilon \in (0,1)$ we have 
\begin{equation}
\label{eqn:int_1}
\lambda_r + \epsilon  <\sum_{\bm z \in \mathcal Z}\mu^{\pi'}(\bm z)\sum_{(\bm{\sigma} ,\bm{n})\in\mathcal{S}}p(\bm{\sigma},\bm{n} | \bm z)   n_r \gamma_r
= \E_{z \sim \mu^{{\pi'}}}[\gamma_r{\bar{ n}_r}]\, , \qquad \forall r  \in \mathcal R \,  .
\end{equation}
Using~\eqref{eqn:int_1} in~\eqref{eqn:lyapunov_drift} we can write 
\begin{align}
    \frac{dL(\bar{\bm Q}(t))}{dt}
    &\leq
    \sum_{r \in \mathcal R}\bar{ Q}_r(t)  \E_{z \sim \mu^{{\pi'}}}[\gamma_r{\bar{ n}_r}]
    -\sum_{r\in \mathcal{R}} \bar{Q}_r(t) \bar{D}'_r(t)
    -\epsilon
    \sum_{r \in \mathcal R}\bar{ Q}_r(t).
    \label{eqn:lyapunov_drift_2}
\end{align}
Note that the term $\sum_{r\in \mathcal{R}}\sum_{\bm z \in \mathcal Z}\mu^{\pi'}(\bm z)\sum_{(\bm{\sigma} ,\bm{n})\in\mathcal{S}}p(\bm{\sigma},\bm{n} | \bm z)   n_r \gamma_r \bar{Q}_r(t)$ corresponds to the average reward under the policy $\pi'$. 
Integrating and applying \eqref{eqn:fl_3} and then~\eqref{eqn:int_1} we see that 
\begin{align*}
L(\bar{\bm Q}(t))
-
L(\bar{\bm Q}(s))
&
\leq 
\int_{s}^t 
    \sum_{r \in \mathcal R}\bar{ Q}_r(u)  \E_{z \sim \mu^{{\pi'}}}[\gamma_r{\bar{ n}_r}]
du
    -
\int_{s}^t 
    \sum_{r\in \mathcal{R}} \bar{Q}_r(u) d\bar{D}_r(u)
    -
\int_{s}^t 
    \epsilon
    \sum_{r \in \mathcal R}\bar{ Q}_r(u)   
du
\\
&
\leq 
    -
\int_{s}^t 
    \epsilon
    \sum_{r \in \mathcal R}\bar{ Q}_r(u)   du
\leq -\frac{\epsilon }{\sqrt{|\mathcal R|}}
\int_s^t L(Q(u))^{\frac{1}{2}}du \, .
\end{align*}
The last inequality follows using the bound: $\norm{\bar{\bm Q}(u)}_1\geq \frac{\norm{\bar{\bm Q}(u)}_2}{\sqrt{|\mathcal{R}|}}$. From this, we see that at any point of differentiability
\begin{equation}
\label{eqn:lyapunov_drift_3}
\frac{dL(\bar{\bm Q}(t))}{dt} \leq -\frac{\epsilon }{\sqrt{|\mathcal R|}} \brac{L(\bar{\bm Q}(t))}^{1/2}.   
\end{equation} 
%
From~\eqref{eqn:lyapunov_drift_3} it can be observed that if $L(\bar{\bm Q}(T)) = 0$ at any differentiable point $ T$, then $L(\bar{\bm Q}(t)) = 0 $ for all $t \geq T $. On the other hand, while $L(\bar{\bm Q}(t)) > 0$, we have from~\eqref{eqn:lyapunov_drift_3} 
\begin{align*}
\brac{L(\bar{\bm Q}(t))}^{1/2}-\brac{L(\bar{\bm Q}(0))}^{1/2}=\frac{1}{2}\int_0^t \brac{L(\bar{\bm Q}(t))}^{-1/2} \frac{dL(\bar{\bm Q}(t))}{dt} dt \leq -\frac{\epsilon }{2\sqrt{|\mathcal R|}}t.
\end{align*}
Thus,
$$
L(\bar{\bm Q}(t)) \leq \Big( {\brac{L(\bar{\bm Q}(0))}^{1/2} -\frac{\epsilon }{2\sqrt{|\mathcal R|}}t} \Big)_+^2.
$$
Moreover, the function $L(\bar{\bm Q}(t))$ is continuous and non-increasing. Hence, for $\norm{\bar{\bm Q}(0)}_1\neq0$, $L(\bar{\bm Q}(t))=0$ for all $t\geq T$ where 
$$
T
=\frac{
        2\sqrt{|\mathcal R|}\brac{L(\bar{\bm Q}(0))}^{1/2}
    }{
        \epsilon
        },
$$
and we have $\bar{Q}_r(t)=0$ for $r\in \mathcal{R}$, for all $t\geq T$. This completes the proof of Theorem~\ref{thm:fluid_stability}. \hfill $\blacksquare$

\section{Conclusion}
This work proposes an optimal scheduling scheme for LLEs with finite lifetimes in a quantum switch to serve incoming entanglement requests. We analyze a general graph topology, which is significantly different from the simplified topologies explored in prior works.  

A key contribution is a novel, asymptotically optimal scheduling policy called the ARE policy. The ARE policy optimizes the utilization of link-level entanglements to serve incoming requests efficiently. Furthermore, the paper makes significant technical contributions by developing a novel method to establish the fluid limit using two-time scale separation in a general quantum switch. Our work has substantive implications for advances in quantum networking. In future work, we will study entanglement purification protocols that require analysis of two-sided queues. A future direction is to establish guarantees for the mean delay of the ARE policy.

 \noindent \textbf{Acknowledgement.} This research was funded by the EPSRC funded INFORMED-AI project EP/Y028732/1. We are also grateful to ongoing discussions with Nu Quantum on quantum networking.

\bibliographystyle{IEEEtran}
\bibliography{ref}

\begin{thebibliography}{10}
\providecommand{\url}[1]{#1}
\csname url@samestyle\endcsname
\providecommand{\newblock}{\relax}
\providecommand{\bibinfo}[2]{#2}
\providecommand{\BIBentrySTDinterwordspacing}{\spaceskip=0pt\relax}
\providecommand{\BIBentryALTinterwordstretchfactor}{4}
\providecommand{\BIBentryALTinterwordspacing}{\spaceskip=\fontdimen2\font plus
\BIBentryALTinterwordstretchfactor\fontdimen3\font minus
  \fontdimen4\font\relax}
\providecommand{\BIBforeignlanguage}[2]{{%
\expandafter\ifx\csname l@#1\endcsname\relax
\typeout{** WARNING: IEEEtran.bst: No hyphenation pattern has been}%
\typeout{** loaded for the language `#1'. Using the pattern for}%
\typeout{** the default language instead.}%
\else
\language=\csname l@#1\endcsname
\fi
#2}}
\providecommand{\BIBdecl}{\relax}
\BIBdecl

\bibitem{wehner_qinternet}
\BIBentryALTinterwordspacing
S.~Wehner, D.~Elkouss, and R.~Hanson, ``Quantum internet: A vision for the road
  ahead,'' \emph{Science}, vol. 362, no. 6412, p. eaam9288, 2018. [Online].
  Available: \url{https://www.science.org/doi/abs/10.1126/science.aam9288}
\BIBentrySTDinterwordspacing

\bibitem{calefi_qi}
\BIBentryALTinterwordspacing
M.~Caleffi, A.~S. Cacciapuoti, and G.~Bianchi, ``Quantum internet: from
  communication to distributed computing!'' in \emph{Proceedings of the 5th ACM
  International Conference on Nanoscale Computing and Communication}, ser.
  NANOCOM '18.\hskip 1em plus 0.5em minus 0.4em\relax New York, NY, USA:
  Association for Computing Machinery, 2018. [Online]. Available:
  \url{https://doi.org/10.1145/3233188.3233224}
\BIBentrySTDinterwordspacing

\bibitem{HK_LO_qinternet}
\BIBentryALTinterwordspacing
K.~Azuma, S.~E. Economou, D.~Elkouss, P.~Hilaire, L.~Jiang, H.-K. Lo, and
  I.~Tzitrin, ``Quantum repeaters: From quantum networks to the quantum
  internet,'' \emph{Rev. Mod. Phys.}, vol.~95, p. 045006, Dec 2023. [Online].
  Available: \url{https://link.aps.org/doi/10.1103/RevModPhys.95.045006}
\BIBentrySTDinterwordspacing

\bibitem{yin2020entanglement}
J.~Yin, Y.-H. Li, S.-K. Liao, M.~Yang, Y.~Cao, L.~Zhang, J.-G. Ren, W.-Q. Cai,
  W.-Y. Liu, S.-L. Li \emph{et~al.}, ``Entanglement-based secure quantum
  cryptography over 1,120 kilometres,'' \emph{Nature}, vol. 582, no. 7813, pp.
  501--505, 2020.

\bibitem{europeancommission2021}
{European Commission}, ``Euroqci: Europe's quantum communication
  infrastructure,'' European Commission publication, 2021,
  \url{https://digital-strategy.ec.europa.eu/en/policies/european-quantum-communication-infrastructure-euroqci}.

\bibitem{toshiba2021quantum}
{Toshiba Europe Press Release}, ``Toshiba and bt launch uk’s first commercial
  quantum secured metro network,'' Press release, October 2021,
  \url{https://www.global.toshiba/ww/news/corporate/2022/04/news-20220427-01.html}.

\bibitem{day2024entangled}
C.~Day, ``Entangled photons maintained under new york streets,''
  \emph{Physics}, vol.~17, p. 125, 2024.

\bibitem{liu2024creation}
J.-L. Liu, X.-Y. Luo, Y.~Yu, C.-Y. Wang, B.~Wang, Y.~Hu, J.~Li, M.-Y. Zheng,
  B.~Yao, Z.~Yan \emph{et~al.}, ``Creation of memory--memory entanglement in a
  metropolitan quantum network,'' \emph{Nature}, vol. 629, no. 8012, pp.
  579--585, 2024.

\bibitem{castelvecchi2024quantum}
D.~Castelvecchi, ``'quantum internet'demonstration in cities is most advanced
  yet.'' \emph{Nature}, 2024.

\bibitem{armstrong2012programmable}
S.~Armstrong, J.-F. Morizur, J.~Janousek, B.~Hage, N.~Treps, P.~K. Lam, and
  H.-A. Bachor, ``Programmable multimode quantum networks,'' \emph{Nature
  communications}, vol.~3, no.~1, p. 1026, 2012.

\bibitem{calefi_qdc}
A.~S. Cacciapuoti, M.~Caleffi, F.~Tafuri, F.~S. Cataliotti, S.~Gherardini, and
  G.~Bianchi, ``Quantum internet: Networking challenges in distributed quantum
  computing,'' \emph{IEEE Network}, vol.~34, no.~1, pp. 137--143, 2020.

\bibitem{quantum_chemistry}
K.~Bourzac, ``4 tough chemistry problems that quantum computers will solve
  [news],'' \emph{IEEE Spectrum}, vol.~54, pp. 7--9, 11 2017.

\bibitem{Cooling_systems}
\BIBentryALTinterwordspacing
L.~Bassman~Oftelie, A.~De~Pasquale, and M.~Campisi, ``Dynamic cooling on
  contemporary quantum computers,'' \emph{PRX Quantum}, vol.~5, p. 030309, Jul
  2024. [Online]. Available:
  \url{https://link.aps.org/doi/10.1103/PRXQuantum.5.030309}
\BIBentrySTDinterwordspacing

\bibitem{nielsen2001quantum}
M.~A. Nielsen and I.~L. Chuang, \emph{Quantum computation and quantum
  information}.\hskip 1em plus 0.5em minus 0.4em\relax Cambridge university
  press Cambridge, 2001, vol.~2.

\bibitem{rajak2023quantum}
A.~Rajak, S.~Suzuki, A.~Dutta, and B.~K. Chakrabarti, ``Quantum annealing: An
  overview,'' \emph{Philosophical Transactions of the Royal Society A}, vol.
  381, no. 2241, p. 20210417, 2023.

\bibitem{briegel2009measurement}
H.~J. Briegel, D.~E. Browne, W.~D{\"u}r, R.~Raussendorf, and M.~Van~den Nest,
  ``Measurement-based quantum computation,'' \emph{Nature Physics}, vol.~5,
  no.~1, pp. 19--26, 2009.

\bibitem{bartolucci2023fusion}
S.~Bartolucci, P.~Birchall, H.~Bombin, H.~Cable, C.~Dawson, M.~Gimeno-Segovia,
  E.~Johnston, K.~Kieling, N.~Nickerson, M.~Pant \emph{et~al.}, ``Fusion-based
  quantum computation,'' \emph{Nature Communications}, vol.~14, no.~1, p. 912,
  2023.

\bibitem{bartolucci2021switch}
S.~Bartolucci, P.~Birchall, D.~Bonneau, H.~Cable, M.~Gimeno-Segovia,
  K.~Kieling, N.~Nickerson, T.~Rudolph, and C.~Sparrow, ``Switch networks for
  photonic fusion-based quantum computing,'' \emph{arXiv preprint
  arXiv:2109.13760}, 2021.

\bibitem{ramaswami2009optical}
R.~Ramaswami, K.~Sivarajan, and G.~Sasaki, \emph{Optical networks: a practical
  perspective}.\hskip 1em plus 0.5em minus 0.4em\relax Morgan Kaufmann, 2009.

\bibitem{georgiadis2006resource}
L.~Georgiadis, M.~J. Neely, L.~Tassiulas \emph{et~al.}, ``Resource allocation
  and cross-layer control in wireless networks,'' \emph{Foundations and
  Trends{\textregistered} in Networking}, vol.~1, no.~1, pp. 1--144, 2006.

\bibitem{tassiulas1990stability}
L.~Tassiulas and A.~Ephremides, ``Stability properties of constrained queueing
  systems and scheduling policies for maximum throughput in multihop radio
  networks,'' in \emph{29th IEEE Conference on Decision and Control}.\hskip 1em
  plus 0.5em minus 0.4em\relax IEEE, 1990, pp. 2130--2132.

\bibitem{mckeown1999achieving}
N.~McKeown, A.~Mekkittikul, V.~Anantharam, and J.~Walrand, ``Achieving 100\%
  throughput in an input-queued switch,'' \emph{IEEE Transactions on
  Communications}, vol.~47, no.~8, pp. 1260--1267, 1999.

\bibitem{mckeown1999islip}
N.~McKeown, ``The islip scheduling algorithm for input-queued switches,''
  \emph{IEEE/ACM transactions on networking}, vol.~7, no.~2, pp. 188--201,
  1999.

\bibitem{zubeldia2022matching}
M.~Zubeldia, P.~R. Jhunjhunwala, and S.~T. Maguluri, ``Matching queues with
  abandonments in quantum switches: Stability and throughput analysis,''
  \emph{arXiv preprint arXiv:2209.12324}, 2022.

\bibitem{Longbo}
\BIBentryALTinterwordspacing
J.~Huang and L.~Huang, ``Learning-based optimal quantum switch scheduling,''
  \emph{SIGMETRICS Perform. Eval. Rev.}, vol.~51, no.~2, p. 75–77, Oct. 2023.
  [Online]. Available: \url{https://doi.org/10.1145/3626570.3626597}
\BIBentrySTDinterwordspacing

\bibitem{valls2024brief}
V.~Valls, P.~Promponas, and L.~Tassiulas, ``A brief introduction to quantum
  network control,'' \emph{arXiv preprint arXiv:2407.19899}, 2024.

\bibitem{promponas2023full}
P.~Promponas, V.~Valls, and L.~Tassiulas, ``Full exploitation of limited memory
  in quantum entanglement switching,'' in \emph{2023 IFIP Networking Conference
  (IFIP Networking)}.\hskip 1em plus 0.5em minus 0.4em\relax IEEE, 2023, pp.
  1--9.

\bibitem{collins2005quantum}
D.~Collins, N.~Gisin, and H.~De~Riedmatten*, ``Quantum relays for long distance
  quantum cryptography,'' \emph{Journal of Modern Optics}, vol.~52, no.~5, pp.
  735--753, 2005.

\bibitem{mandil2023quantum}
R.~Mandil, S.~DiAdamo, B.~Qi, and A.~Shabani, ``Quantum key distribution in a
  packet-switched network,'' \emph{npj Quantum Information}, vol.~9, no.~1,
  p.~85, 2023.

\bibitem{palacios2018introduction}
C.~Palacios-Berraquero, ``Introduction: 2d-based quantum technologies,''
  \emph{Quantum Confined Excitons in 2-Dimensional Materials}, pp. 1--30, 2018.

\bibitem{Nain_switch}
\BIBentryALTinterwordspacing
P.~Nain, G.~Vardoyan, S.~Guha, and D.~Towsley, ``On the analysis of a
  multipartite entanglement distribution switch,'' \emph{Proc. ACM Meas. Anal.
  Comput. Syst.}, vol.~4, no.~2, Jun. 2020. [Online]. Available:
  \url{https://doi.org/10.1145/3392141}
\BIBentrySTDinterwordspacing

\bibitem{Thiru_switch}
\BIBentryALTinterwordspacing
T.~Vasantam and D.~Towsley, ``{A throughput optimal scheduling policy for a
  quantum switch},'' in \emph{Quantum Computing, Communication, and Simulation
  II}, P.~R. Hemmer and A.~L. Migdall, Eds., vol. 12015, International Society
  for Optics and Photonics.\hskip 1em plus 0.5em minus 0.4em\relax SPIE, 2022,
  p. 1201505. [Online]. Available: \url{https://doi.org/10.1117/12.2616950}
\BIBentrySTDinterwordspacing

\bibitem{10229003}
N.~K. Panigrahy, T.~Vasantam, D.~Towsley, and L.~Tassiulas, ``On the capacity
  region of a quantum switch with entanglement purification,'' in \emph{IEEE
  INFOCOM 2023 - IEEE Conference on Computer Communications}, 2023, pp. 1--10.

\bibitem{Wenhan_switch}
W.~Dai, A.~Rinaldi, and D.~Towsley, ``The capacity region of entanglement
  switching: Stability and zero latency,'' in \emph{2022 IEEE International
  Conference on Quantum Computing and Engineering (QCE)}, 2022, pp. 389--399.

\bibitem{hunt1994large}
P.~Hunt and T.~Kurtz, ``Large loss networks,'' \emph{Stochastic Processes and
  their Applications}, vol.~53, no.~2, pp. 363--378, 1994.

\bibitem{10.1214/14-AAP1057}
\BIBentryALTinterwordspacing
P.~Robert and A.~V{\'e}ber, ``{A stochastic analysis of resource sharing with
  logarithmic weights},'' \emph{The Annals of Applied Probability}, vol.~25,
  no.~5, pp. 2626 -- 2670, 2015. [Online]. Available:
  \url{https://doi.org/10.1214/14-AAP1057}
\BIBentrySTDinterwordspacing

\bibitem{yasodharan2022large}
S.~Yasodharan and R.~Sundaresan, ``Large deviations of mean-field interacting
  particle systems in a fast varying environment,'' \emph{The Annals of Applied
  Probability}, vol.~32, no.~3, pp. 1666--1704, 2022.

\bibitem{10.1145/3152042.3152052}
\BIBentryALTinterwordspacing
G.~Goel, N.~Chen, and A.~Wierman, ``Thinking fast and slow: Optimization
  decomposition across timescales,'' \emph{SIGMETRICS Perform. Eval. Rev.},
  vol.~45, no.~2, p. 27–29, Oct. 2017. [Online]. Available:
  \url{https://doi.org/10.1145/3152042.3152052}
\BIBentrySTDinterwordspacing

\bibitem{bennett1993teleporting}
C.~H. Bennett, G.~Brassard, C.~Cr{\'e}peau, R.~Jozsa, A.~Peres, and W.~K.
  Wootters, ``Teleporting an unknown quantum state via dual classical and
  einstein-podolsky-rosen channels,'' \emph{Physical review letters}, vol.~70,
  no.~13, p. 1895, 1993.

\bibitem{nain2020analysis}
P.~Nain, G.~Vardoyan, S.~Guha, and D.~Towsley, ``On the analysis of a
  multipartite entanglement distribution switch,'' \emph{Proceedings of the ACM
  on Measurement and Analysis of Computing Systems}, vol.~4, no.~2, pp. 1--39,
  2020.

\bibitem{NISQ1}
\BIBentryALTinterwordspacing
J.~W.~Z. Lau, K.~H. Lim, H.~Shrotriya, and L.~C. Kwek, ``Nisq computing: where
  are we and where do we go?'' \emph{AAPPS Bulletin}, vol.~32, pp. 1--30, 2022.
  [Online]. Available: \url{https://api.semanticscholar.org/CorpusID:252538719}
\BIBentrySTDinterwordspacing

\bibitem{NISQ2}
S.~Brandhofer, S.~Devitt, T.~Wellens, and I.~Polian, ``Special session: Noisy
  intermediate-scale quantum (nisq) computers—how they work, how they fail,
  how to test them?'' in \emph{2021 IEEE 39th VLSI Test Symposium (VTS)}, 2021,
  pp. 1--10.

\bibitem{quantum_memory}
Y.~Wang, M.~Um, Z.~Junhua, S.~An, M.~Lyu, J.-n. Zhang, L.~Duan, D.~Yum, and
  K.~Kim, ``Single-qubit quantum memory exceeding $10$-minute coherence time,''
  \emph{Nature Photonics}, vol.~11, 10 2017.

\bibitem{Fittipaldi}
\BIBentryALTinterwordspacing
P.~Fittipaldi, A.~Giovanidis, and F.~Grosshans, ``{ A Linear Algebraic
  Framework for Dynamic Scheduling Over Memory-Equipped Quantum Networks },''
  \emph{IEEE Transactions on Quantum Engineering}, vol.~5, no.~01, pp. 1--18,
  Jan. 2024. [Online]. Available:
  \url{https://doi.ieeecomputersociety.org/10.1109/TQE.2023.3341151}
\BIBentrySTDinterwordspacing

\bibitem{andrews2004scheduling}
M.~Andrews, K.~Kumaran, K.~Ramanan, A.~Stolyar, R.~Vijayakumar, and P.~Whiting,
  ``Scheduling in a queuing system with asynchronously varying service rates,''
  \emph{Probability in the Engineering and Informational Sciences}, vol.~18,
  no.~2, pp. 191--217, 2004.

\bibitem{bonald2001impact}
T.~Bonald and L.~Massouli{\'e}, ``Impact of fairness on internet performance,''
  in \emph{Proceedings of the 2001 ACM SIGMETRICS international conference on
  Measurement and modeling of computer systems}, 2001, pp. 82--91.

\bibitem{puterman2014markov}
M.~L. Puterman, \emph{Markov decision processes: discrete stochastic dynamic
  programming}.\hskip 1em plus 0.5em minus 0.4em\relax John Wiley \& Sons,
  2014.

\bibitem{bertsekas2011dynamic}
D.~P. Bertsekas, ``Dynamic programming and optimal control: Volume ii,''
  \emph{Belmont, MA: Athena Scientific}, vol.~1, 2011.

\bibitem{billingsley2013convergence}
P.~Billingsley, \emph{Convergence of probability measures}.\hskip 1em plus
  0.5em minus 0.4em\relax John Wiley \& Sons, 2013.

\bibitem{bertsimas1998geometric}
D.~Bertsimas, D.~Gamarnik, and J.~N. Tsitsiklis, ``Geometric bounds for
  stationary distributions of infinite markov chains via lyapunov functions,''
  1998.

\bibitem{robert2013stochastic}
P.~Robert, \emph{Stochastic networks and queues}.\hskip 1em plus 0.5em minus
  0.4em\relax Springer Science \& Business Media, 2013, vol.~52.

\bibitem{dudley2002real}
R.~Dudley, \emph{Real analysis and probability}.\hskip 1em plus 0.5em minus
  0.4em\relax Cambridge University Press, 2002.

\bibitem{dai1995positive}
J.~G. Dai, ``On positive harris recurrence of multiclass queueing networks: a
  unified approach via fluid limit models,'' \emph{The Annals of Applied
  Probability}, vol.~5, no.~1, pp. 49--77, 1995.

\bibitem{bramson2008stability}
M.~Bramson, \emph{Stability of queueing networks}.\hskip 1em plus 0.5em minus
  0.4em\relax Springer, 2008.

\bibitem{dai1995stability}
J.~Dai, ``Stability of open multiclass queueing networks via fluid models,''
  \emph{IMA Volumes in Mathematics and Its Applications}, vol.~71, pp. 71--71,
  1995.

\bibitem{qu2020finite}
G.~Qu and A.~Wierman, ``Finite-time analysis of asynchronous stochastic
  approximation and $ q $-learning,'' in \emph{Conference on Learning
  Theory}.\hskip 1em plus 0.5em minus 0.4em\relax PMLR, 2020, pp. 3185--3205.

\bibitem{hoeffding1994probability}
W.~Hoeffding, ``Probability inequalities for sums of bounded random
  variables,'' \emph{The collected works of Wassily Hoeffding}, pp. 409--426,
  1994.

\bibitem{seneta1988perturbation}
E.~Seneta, ``Perturbation of the stationary distribution measured by ergodicity
  coefficients,'' \emph{Advances in Applied Probability}, vol.~20, no.~1, pp.
  228--230, 1988.

\bibitem{seneta2006non}
------, \emph{Non-negative matrices and Markov chains}.\hskip 1em plus 0.5em
  minus 0.4em\relax Springer Science \& Business Media, 2006.

\bibitem{dobrushin1956central}
R.~L. Dobrushin, ``Central limit theorem for nonstationary markov chains. i,''
  \emph{Theory of Probability \& Its Applications}, vol.~1, no.~1, pp. 65--80,
  1956.

\end{thebibliography}

\appendix
\section{Proof of Theorem \ref{thrm:stabilityRegion}}
\label{append:stabilityregion}

We now restate and prove Theorem \ref{thrm:stabilityRegion}.

\ThrmStabilityRegion*

\begin{proof}
We first prove \eqref{eq:Cbar} holds. To do this, we take any stabilizable policy. We then use the long-run service process of that policy to design an LLE scheduling scheme with the correct stationary service rate to satisfy \eqref{eq:Cbar}.

First, if $\bm \lambda \in \mathcal C$, then let $\bm X(\infty)=(\bm Z(\infty),\bm Q(\infty))$ denote the stationary distribution of the Markov chain that is positive recurrent under $\bm \lambda$. Also, we let $ d_r(\infty)$ denote the stationary number type $r$ requests that depart. Thus we have for $r \in \mathcal R$,
\begin{align}
 \lambda_r &\leq\mathbb{E}\sbrac{ d_r(\infty)}\notag \\
&=\sum_{\bm z,\bm q}\mb{P}(\bm X(\infty)=(\bm z,\bm q))\sum_{(\bm{\sigma},\bm{n})\in\mathcal{S}}p(\bm{\sigma},\bm{n} | \bm z, \bm q)  n_r \gamma_r \, . \label{eq:expand}
\end{align}
Above, \eqref{eq:expand} the expanded expression for the stationary departure rate.
Here $p(\bm{\sigma},\bm{n} | \bm z, \bm q) $ denotes the stationary probability the schedule $(\bm{\sigma},\bm{n})$ is selected given the system state $(\bm z, \bm q)$. 

Conditioning on $\bm Z(\infty)$, we can further write
\begin{align}
 \lambda_r
&\leq\sum_{\bm z}
    \mb{P}(\bm Z(\infty)=\bm z) 
    \sum_{\bm q}
        \mb{P}(\bm Q(\infty)=\bm q | \bm Z(\infty) =  \bm z)
    \sum_{(\bm{\sigma},\bm{n})\in\mathcal{S}}
            p(\bm{\sigma},\bm{n} | \bm z, \bm q)
             n_r \gamma_r ,
            \notag \\
 & =\sum_{\bm z}\mb{P}(\bm Z(\infty)=\bm z)\sum_{(\bm{\sigma} ,\bm{n})\in\mathcal{S}}p(\bm{\sigma},\bm{n} | \bm z)   n_r \gamma_r \label{eq:lle_stationary}
\end{align}
where 
\begin{equation}\label{eq:pidef}
p(\bm{\sigma} ,\bm{n} | \bm z)
:=
\sum_{\bm q}
    \mb{P}(\bm Q(\infty)=\bm q\mid \bm Z(\infty)=\bm z)
    p(\bm{\sigma} ,\bm{n} | \bm z, \bm q) 
\end{equation}
 denotes the probability that schedule $(\bm{\sigma} ,\bm{n})$ is selected in stationary regime given that $\bm Z(\infty)=\bm{z}$. 

We can now use it to define an entanglement matching policy that is agnostic to request queues. In particular, we define the request agnostic policy $\hat \pi$ where whenever the left side queues are in state $\bm z$, we choose the schedule $(\bm \sigma, \bm n)$ with probability $p(\bm{\sigma},\bm{n} | \bm z)$ as given by \eqref{eq:pidef}. We let $\hat{\bm Z}(\infty)$ denote the stationary number of LLEs under this policy. 

We now show that $\hat{\bm Z}(\infty)$ and $\bm Z(\infty)$ are equal in distribution.\footnote{For a standard switching model these are immediately equal as they are both independent of the queue size process $\bm Q(\infty)$; however, in our case this property that must be verified. The following calculations are not standard in prior analysis on Maximal Stability.} We prove this by verifying both distributions satisfy identical balance equations.

First the balance equations for $\hat{\bm Z}(\infty)$  are
\begin{equation}
\mathbb P (\hat{\bm Z}(\infty) = \bm z) 
=
\sum_{\bm z'} \sum_{(\bm{\sigma} ,\bm{n})\in\mathcal{S}} \mathbb P 
( \bm Z(1) = \bm z | \bm Z(0) =\bm z', \bm n(0) = \bm n, \bm \sigma(0)= \bm \sigma  ) p(\bm{\sigma} ,\bm{n} | \bm z')
\mathbb P( \hat {\bm Z}(\infty) = \bm z' ) \, .
\label{eq:hatbalance}    
\end{equation}
Here, $\bm z'$ indicates the initial state of the stationary distribution, and $\bm z$ represents the next state reached after one transition.
Next, the balance equations for $\bm X(\infty) = (\bm Z(\infty), \bm Q(\infty))$ are 
:
\begin{align}
    &
    \mathbb P (  \bm Z( \infty) = \bm z, \bm Q(\infty) = \bm q )
    \notag
\\
&
= 
    \sum_{\bm z' \bm q'} \sum_{(\bm{\sigma} ,\bm{n})\in\mathcal{S}} 
    \Big[
    \mathbb P ( \bm Z(1) =\bm z  , \bm Q(1) = \bm q  | \bm Z(0) =\bm z ' , \bm Q(0) = \bm q', \bm n(0)= \bm n, \bm \sigma(0)= \bm \sigma )
    \notag
    \\
    &\qquad \qquad \qquad \qquad
    \times 
    p(\bm{\sigma} ,\bm{n} | \bm z', \bm q')
    \notag
    \\
    &\qquad \qquad \qquad  \qquad\qquad
    \times
    \mathbb P (\bm Z(\infty) =\bm z ' , \bm Q(\infty ) = \bm q' )  
    \Big]\, .
     \label{eq:Zlong0}
\end{align}
Thus, summing over $\bm q$ gives
\begin{align}
  &
  \mathbb P (  \bm Z( \infty) = \bm z)
  \notag
  \\
&
  = \sum_{\bm q} \mathbb P (  \bm Z( \infty) = \bm z, \bm Q(\infty) = \bm q ) 
  \notag
  \\
  &
= 
    \sum_{\bm q\, \bm z' \bm q'} \sum_{(\bm{\sigma} ,\bm{n})\in\mathcal{S}} 
    \Big[ 
    \mathbb P ( \bm Z(1) =\bm z  , \bm Q(1) = \bm q  | \bm Z(0) =\bm z ' , \bm Q(0) = \bm q' , \bm n(0)= \bm n, , \bm \sigma(0)= \bm \sigma)  
    \notag
    \\
    &
    \qquad \qquad \qquad \qquad 
    \times
    p(\bm{\sigma} ,\bm{n} | \bm z', \bm q')
    \notag
    \\
    &\qquad \qquad \qquad \qquad\quad \times
    \mathbb P (\bm Z(\infty) =\bm z ' , \bm Q(\infty ) = \bm q' )  \Big]
    \label{eq:Zlong1}
\\
&
=
    \sum_{\, \bm z' \bm q'} \sum_{(\bm{\sigma} ,\bm{n})\in\mathcal{S}} 
   \Big[ 
    \mathbb P ( \bm Z(1) =\bm z  | \bm Z(0) =\bm z ' , \bm Q(0) = \bm q' , \bm n(0)= \bm n, \bm \sigma(0)= \bm \sigma)  
    \notag
    \\
    &
    \qquad \qquad \qquad \qquad 
    \times
    p(\bm{\sigma} ,\bm{n} | \bm z', \bm q')
    \notag
    \\
    &\qquad \qquad \qquad \qquad\quad \times
    \mathbb P (\bm Z(\infty) =\bm z ' , \bm Q(\infty ) = \bm q' )  \Big]
    \label{eq:Zlong2}
\\
&
=
    \sum_{\, \bm z' \bm q'} \sum_{(\bm{\sigma} ,\bm{n})\in\mathcal{S}} 
   \Big[ 
    \mathbb P ( \bm Z(1) =\bm z  | \bm Z(0) =\bm z ' ,  \bm n(0)= \bm n, \bm \sigma(0)= \bm \sigma) 
    \notag
    \\
    &
    \qquad \qquad \qquad \qquad 
    \times
    p(\bm{\sigma} ,\bm{n} | \bm z', \bm q')
    \notag
    \\
    &\qquad \qquad \qquad \qquad\quad \times
    \mathbb P ( \bm Q(\infty ) = \bm q' 
 | \bm Z(\infty) =\bm z '  ) 
     \mathbb P (\bm Z(\infty) =\bm z ' ) \Big]  
     \label{eq:Zlong3}
\\
&
=
    \sum_{\, \bm z' }
    \sum_{(\bm{\sigma} ,\bm{n})\in\mathcal{S}} 
    \mathbb P ( \bm Z(1) =\bm z  | \bm Z(0) =\bm z ' ,  \bm n(0)= \bm n, \bm \sigma(0)= \bm \sigma) 
    \notag
    \\
    &
    \qquad \qquad \qquad  
    \times
    \Big[  
    \sum_{\, \bm q' }
   p(\bm{\sigma} ,\bm{n} | \bm z', \bm q')
    \mathbb P ( \bm Q(\infty ) = \bm q' 
 | \bm Z(\infty) =\bm z '  ) 
 \Big]
     \mathbb P (\bm Z(\infty) =\bm z ' )   
      \label{eq:Zlong4}
\\
&
=
    \sum_{\, \bm z' }
    \sum_{(\bm{\sigma} ,\bm{n})\in\mathcal{S}} 
    \mathbb P ( \bm Z(1) =\bm z  | \bm Z(0) =\bm z ' ,  \bm n(0)= \bm n, \bm \sigma(0)= \bm \sigma)
    p(\bm{\sigma} ,\bm{n} | \bm z')
         \mathbb P (\bm Z(\infty) =\bm z ' )   \, .
          \label{eq:Zlong5}
\end{align}
Above in \eqref{eq:Zlong1}, we apply \eqref{eq:Zlong0}. In \eqref{eq:Zlong2} we sum over $\bm q$. In \eqref{eq:Zlong3}, we note that the transition from $\bm z'$ to $\bm z$ does not depend on $\bm q'$ once we condition on the schedule $\bm n$, and we also condition on $\bm z'$. In \eqref{eq:Zlong4} we notice we can bring sum over $\bm q'$ inside. In \eqref{eq:Zlong5} we note by our definition \eqref{eq:pidef} that this inner sum is $p(\bm{\sigma} ,\bm{n} | \bm z')$. From \eqref{eq:Zlong5} and \eqref{eq:hatbalance} we see that both $\bm Z(\infty)$, $\hat {\bm Z}(\infty)$ satisfy the same balance equation (on the same irreducible set of states). Thus by uniqueness of stationary distributions $\bm Z(\infty)$, $\hat {\bm Z}(\infty)$ are equal in distribution.

We can now see that \eqref{eq:Cbar} holds. In particular, we let $\mu (\bm z) = \mathbb P(\hat {\bm Z}(\infty) = \bm z) = \mathbb P(\bm Z(\infty) = \bm z)$ then, from \eqref{eq:expand}, we see that our request agnostic policy $\hat \pi$ is such that 
\[
\lambda_r
 \leq \sum_{\bm z}\mu(\bm z)\sum_{(\bm{\sigma} ,\bm{n})\in\mathcal{S}}p(\bm{\sigma},\bm{n} | \bm z)   n_r \gamma_r  \,. 
\]
This verifies \eqref{eq:Cbar} and completes the first part of the theorem.

For the second part, we apply Foster's Lemma.\footnote{Again, there are some complications when compared to more traditional maximal stability proofs due to the random evolution of the service process.} Consider an agnostic policy such that 
\[
\lambda_r
 < \sum_{\bm z}\mu(\bm z)\sum_{(\bm{\sigma} ,\bm{n})\in\mathcal{S}}p(\bm{\sigma},\bm{n} | \bm z)   n_r \gamma_r  \, 
\]
holds.
Notice that if the LLE process is stationary, then we have for all $\bm q$ sufficiently large
\begin{align*}
& \mathbb E \Big[ \sum_r Q_r(t+1) -\sum_r Q_r(t) | \bm Q(t)=\bm q, \bm Z(t)\sim \mu \Big]
\\
& = 
\sum_r \Big( \lambda_r  - \sum_{\bm z}\mu(\bm z)\sum_{(\bm{\sigma} ,\bm{n})\in\mathcal{S}}p(\bm{\sigma},\bm{n} | \bm z)   n_r \gamma_r  \Big) <0 \, .
\end{align*}
Thus, in principle, if it were not for Markov evolution of the $\bm Z(t)$ process, then we could directly apply Foster's Lemma. In order to deal with this issue, we simply need to give the process sufficient time to approach equilibrium and then we can apply Lyapunov arguments. We start by proving stability for one queue, since $(\bm Z(t),  Q_r(t))$ is a Markov chain for agnostic policies. We then use our queue size bound to prove stability for $(\bm Z(t),  \bm Q(t))$. We apply the Lyapunov argument of Bertsimas et al. \cite{bertsimas1998geometric}. This is the plan for the remainder of the proof. \footnote{We note that an alternative approach is to Apply Foster-Lyapunov over renewal cycles of the process $\bm Z$ and apply Theorem 8.13 of Robert \cite{robert2013stochastic}. Another approach would be to consider a quadratic Lyapunov function and then apply a similar argument to \cite{georgiadis2006resource}.}

We let 
\[
\epsilon_r :=  \Big( \sum_{\bm z}\mu(\bm z)\sum_{(\bm{\sigma} ,\bm{n})\in\mathcal{S}}p(\bm{\sigma},\bm{n} | \bm z)   n_r \gamma_r \Big)   -\lambda_r  \, .
\]
We note that under this policy the lefthand process $\bm Z(t)$ is an ergodic Markov chain thus there exists a $t_0$ such for all $t\geq t_0$ 
\[
\left|\sum_{\bm z}\mu(\bm z)\sum_{(\bm{\sigma} ,\bm{n})\in\mathcal{S}}p(\bm{\sigma},\bm{n} | \bm z)   n_r \gamma_r  -
\sum_{\bm z}\mathbb P (\bm Z(t) = \bm z)\sum_{(\bm{\sigma} ,\bm{n})\in\mathcal{S}}p(\bm{\sigma},\bm{n} | \bm z)   n_r \gamma_r  \right| < \frac{\epsilon_r}{2}\, .
\] 
Thus for $t \geq t_0$, we have 
\[
 \mathbb E [ Q_r(t+1) - Q_r(t)  | Q_r(t)=q_r ] \leq -\frac{\epsilon_r}{2} \, ,
\]
for any $q_r >B$. We can apply Foster-Lyapunov to the above expression to prove that  $(\bm Z(t), Q_r(t))$ is a positive recurrent Markov chain. Further by \cite{bertsimas1998geometric}, we have that $\mathbb E [ Q_r(\infty) ] < \infty$ and thus $\mathbb E [ \sum_r Q_r(\infty) ] < \infty$ and thus we see that the Markov chain $(\bm Z(t), \bm Q(t))$ is positive recurrent.



\end{proof}

\section{Tightness: Proof of Proposition \ref{prop:Tightness}} \label{app:tightness}

We now restate and prove Proposition \ref{prop:Tightness}.

\PropositionTightness*

\begin{proof}
To prove tightness, we show that there exists a measurable set $G$ with $\mathbb{P}(G) = 1$, such that for all $\omega \in G$, any subsequence of $\brac{\bm \lambda^c,\bar{\bm{Q}}^c,\bar{\bm{A}}^c,\bar{\bm{D}}^c, \bar{\bm Z}^c}_{c \in \mathbb{N}}$ under the ARE policy contains a further subsequence that converges uniformly on compact time intervals.

Firstly, since $\bm \lambda^c$ belongs to the compact set $\mathcal C$, we can chose a subsequent along which $\bm \lambda^c$ converges to some value $\bm \lambda$. 
As discussed in Subsection~\ref{subsec:topology}, for each $r \in \mathcal{R}$, the collection $(A_r(t) - A_r(t-1) : t \in \Nats)$ consists of i.i.d. random variables with mean $\lambda_r$. Therefore, by the Functional Strong Law of Large Numbers, on a set $G_1$ with $\mathbb{P}(G_1) = 1$, we have for each $r \in \mathcal{R}$ 
$$\bar{A}^c_r(t) \to \lambda_r t,$$
as $c \to \infty$, with the convergence being uniform on compact intervals. Similarly since $Z_l(t)$ is bounded we have
\[
\bar{Z}_l^c(t) \rightarrow 0
\]
as $c\rightarrow\infty$, with the convergence being uniform on compact intervals.
From the Arzelà-Ascoli Theorem, we know that any sequence of equicontinuous functions $\bar{Y}^c(t)$ on $[0,T]$ for $T > 0$, with $\sup_c\left|\bar{Y}^c(0)\right| < \infty$, has a converging subsequence with respect to the uniform norm. We will now verify that any subsequence of $\brac{\bm \lambda^c,\bar{\bm{Q}}^c,\bar{\bm{A}}^c,\bar{\bm{D}}^c, \bar{\bm Z}^c}_{c \in \mathbb{N}}$ satisfies both conditions for any $\omega \in G_1$.

Since $\bar{A}_r^c(0)$ and $\bar{D}_r^c(0)$ are initially $0$ and $\|\bar{\bm Q}^c(0)\|_1 \leq 1$, the supremum is also bounded. The equicontinuity of the sequence $\bar{A}_r^c$ follows from uniform convergence on compact intervals. Moreover, the fact that the maximum number of requests served in each time slot is bounded above by $B$ implies that $\bar{D}_r^c$ is Lipschitz continuous, which in turn implies the equicontinuity of this sequence. Finally, since $\bar{Q}_r^c$ is the sum of a bounded number of equicontinuous functions, the sequence $\bar{Q}_r^c$ is also equicontinuous. Therefore, as the conditions of the Arzelà-Ascoli Theorem are satisfied for all $\omega \in G_1$, every subsequence of $\brac{\bm \lambda^c,\bar{\bm{Q}}^c,\bar{\bm{A}}^c,\bar{\bm{D}}^c, \bar{\bm Z}^c}_{c \in \mathbb{N}}$ contains a further subsequence that converges uniformly. Moreover, since these sequences of functions are uniformly Lipschitz continuous, their limits must also be Lipschitz continuous.
\end{proof}

\section{Optimality of Limit Points: Proof of Proposition \ref{prop:opt}}\label{sec:proofopt}
\label{append:proofopt}

\OptProp*

\begin{proof}
We quickly note the following technical point before proceeding with the proof: $\bar{D}_r(u)$ is an increasing Lipschitz continuous function, and thus, its derivative exists everywhere except a set of measure zero. Thus we can take the Lebegue integral of the derivative of $\bar{D}'_r(u)$ when it exists. See Dudley \cite{dudley2002real}, for instance, for further details.

    To see why \eqref{eqn:mdp_optimal} holds, we observe that the prelimit summations are Riemann integral approximations for the integrals above. In particular, notice that by \eqref{eq:Dlim}
\begin{align}\label{eq:d1}
    \int_s^t \bar{ Q}_r(u)  \E_{z \sim \mu^{\bar{\pi}}}[\gamma_r{\bar{ n}_r}]du
    &= 
    \lim_{c\rightarrow\infty}\frac{1}{c} \sum_{i=\floor{cs}+1}^{\floor{ct}} \bar{ Q}_r^c\big({{\left\lfloor i \right\rfloor_{\tau(c)} }} \big)  \E_{z \sim \mu^{\bar{\pi}}}[\gamma_r{\bar{ n}_r}]\, ,
\\ \label{eq:d2}
    \int_s^t {\bar Q}_r(u) 
    d \bar{ D}_r(u)
    &= 
    \lim_{c\rightarrow\infty} \frac{1}{c}\sum_{i=\floor{cs}+1}^{\floor{ct}} \bar{ Q}_r^c\big({{\left\lfloor i \right\rfloor_{\tau(c)} }}\big)  \E_{z\sim \mu^{\pi^\star(\bm Q(i\tau(c))}} \sbrac{\gamma_r {n}^\star_r},
\end{align}
where $\left\lfloor i \right\rfloor_{\tau(c)} :=\left\lfloor \frac{i}{\tau(c)} \right\rfloor \frac{\tau(c)}{c}$ is the most recent time index,  as a multiple of $ \tau(c)$. 

Notice by the definition of $\pi^\star(\bm Q(i\tau(c))$ being the optimal average MDP solution c.f. \eqref{eq:AREMDP}, we have for all $i$ 
\begin{equation}\label{eq:d3}
\E_{z\sim \mu^{\pi^\star(\bm Q(i\tau(c))}} \Big[
\sum_{r \in \mathcal R}\bar{ Q}_r^c\big({{\left\lfloor i \right\rfloor_{\tau(c)} }
\big)  \gamma_r {n}^\star_r}
\Big]
\geq 
       \E_{z \sim \mu^{\bar{\pi}}}
       \Big[
     \sum_{r \in \mathcal R}
     \bar{ Q}_r^c\big({{\left\lfloor i \right\rfloor_{\tau(c)} }} \big)\gamma_r{\bar{ n}_r}
     \Big].
\end{equation}

Combining \eqref{eq:d1}, \eqref{eq:d2} and \eqref{eq:d3} we see that  
\begin{align*}
        \int_s^t \sum_{r \in \mathcal R} \bar{ Q}_r(u) 
     \bar{ D}'_r(u) du
    &
    = 
    \lim_{c\rightarrow\infty}\frac{1}{c} \sum_{i=\floor{cs}+1}^{\floor{ct}} \sum_{r \in \mathcal R}\bar{ Q}_r^c\big({{\left\lfloor i \right\rfloor_{\tau(c)} }}\big)  \E_{z\sim \mu^{\pi^\star(\bm Q(i\tau(c))}} \sbrac{\gamma_r {n}^\star_r}
    \\
    &=
    \lim_{c\rightarrow\infty} \frac{1}{c}\sum_{i=\floor{cs}+1}^{\floor{ct}}\sum_{r \in \mathcal R}
    \E_{z\sim \mu^{\pi^\star(\bm Q(i\tau(c))}} \Big[
\sum_{r \in \mathcal R}\bar{ Q}_r^c\big({{\left\lfloor i \right\rfloor_{\tau(c)} }
\big)  \gamma_r {n}^\star_r}
\Big]
\\
&
\geq 
\lim_{c\rightarrow\infty} \frac{1}{c}
\sum_{i=\floor{cs}+1}^{\floor{ct}}
\sum_{r \in \mathcal R}
       \E_{z \sim \mu^{\bar{\pi}}}
       \Big[
     \sum_{r \in \mathcal R}
     \bar{ Q}_r^c\big({{\left\lfloor i \right\rfloor_{\tau(c)} }} \big)\gamma_r{\bar{ n}_r}
     \Big]
     \\
     &
     =
         \lim_{c\rightarrow\infty} \frac{1}{c}\sum_{i=\floor{cs}+1}^{\floor{ct}} 
         \sum_{r \in \mathcal R}
         \bar{ Q}_r^c\big({{\left\lfloor i \right\rfloor_{\tau(c)} }} \big)\cdot  \E_{z \sim \mu^{\bar{\pi}}}[\gamma_r{\bar{ n}_r}]
    =
        \int_s^t \sum_{r \in \mathcal R}\bar{ Q}_r(u)  \E_{z \sim \mu^{\bar{\pi}}}[\gamma_r{\bar{ n}_r}]du.
\end{align*}
Thus we see that~\eqref{eqn:fl_3} hold as required.
\end{proof}

\section{Stochastic Stability: Proof of Theorem~\ref{thm:stochastic_stability}}
\label{append:stochastic_stability}
To prove the positive recurrence of the Markov process $\bar{\bm X}^c$ under the policy $\pi^\star(\tau(c))$, Proposition~\ref{prop:Tightness}, Theorem~\ref{thm:fluid_stability} and general stability results from \cite{dai1995positive, bramson2008stability, dai1995stability} are applied.
The main idea is to leverage the $L_1$ convergence of a subsequence of $\brac{\bar{\bm Q}^c(t)}_{c \in \Nats}$ in conjunction with the multiplicative Foster's Lemma.

Note that if the policies $\pi^\star(\tau(c))$ were not asymptotically throughput optimal then the following converse claim must hold:
\begin{equation}
\label{eq:A}
\parbox{\dimexpr\linewidth-4em}{%
    \strut
        There exists an $\epsilon>0$ such that for all $c_\epsilon\in\mathbb N$ there exists a $c>c_\epsilon$ and an arrival rate vector $\bm \lambda^c \in \mathcal C^\epsilon$ such that the policy $\pi^\star(\tau(c))$ is not positive recurrent. 
    \strut
  }
\end{equation}

We will now argue that this cannot be the case. 

We consider the sequence of arrival rates and queueing networks $\brac{\bm \lambda^c,\bar{\bm{Q}}^c,\bar{\bm{A}}^c,\bar{\bm{D}}^c, \bar{\bm Z}^c}_{c\in \Nats}$ as described in \eqref{eq:A} each operating under the policy $\pi^\star(\tau(c))$.
Since $\mathcal C^\epsilon$ is compact, any limit point of $\bm \lambda^c$  belongs to $\mathcal C^\epsilon$. 
Recall that for a sequence of random variables converges in distribution and is uniformly integrable, convergences in $L_1$.
By Proposition~\ref{prop:Tightness} and Theorem \ref{thm:fluid_convergence}, $\brac{\bm \lambda^c,\bar{\bm{Q}}^c,\bar{\bm{A}}^c,\bar{\bm{D}}^c, \bar{\bm Z}^c}_{c}$ has a subsequence that converges to $\brac{\bm \lambda,\bar{\bm{Q}},\bm \lambda t,\bar{\bm{D}}^c, \bm 0 t }$ where $\bar{\bm Q}$ fluid solution with arrival rate $\bm \lambda \in \mathcal C^\epsilon$. Let the subsequence be $\brac{\bm \lambda^{c_k},\bar{\bm{Q}}^{c_k},\bar{\bm{A}}^{c_k},\bar{\bm{D}}^{c_k}, \bar{\bm Z}^{c_k}}_{k \in \mathbb{N}}$. From the fluid stability Theorem~\ref{thm:fluid_stability} 
notice that $T$ can be chosen uniformly for all $\bm \lambda \in \mathcal C^\epsilon$. This observation will help us prove asymptotic throughput optimality.
, we know that there exits a $T>0$ such that for all $\bm \lambda \in \mathcal C^\epsilon$ $\bar{Q}_r(T)=0$ for $r\in \mathcal{R}$. Therefore, the subsequence $\brac{\bar{\bm Z}^{c_k}(T),\bar{\bm Q}^{c_k}(T)}_{k\in \Nats}$ converges in distribution to $(\bar{\bm Z}(T),\bar{\bm Q}(T))=(\bm 0,\bm 0)$.
Next, we claim that the subsequence $\brac{\bar{\bm Z}^{c_k}(T),\bar{\bm Q}^{c_k}(T)}_{k\in \Nats}$ is uniformly integrable.

It is important to note that the sequence $\brac{\bar{\bm Z}^c(t),\bar{\bm Q}^c(t)}_{c \in \Nats}$ is uniformly integrable for any $t\geq0$. This is because the unscaled LLE process ${\bm Z}^c(t)$ is uniformly bounded and because the queue length process $\bar{\bm Q}^c(t)$ is bounded above by the cumulative request arrival process $\bar{\bm A}^c(t)$ plus $\bar {\bm {Q}}^c(0)$. Moreover, we have $\mathbb{E}[(\bar{\bm A}^c(t))^2]$ is bounded as the increments of $\bar{\bm A}^c(t)$ are assumed to have bounded variance. Hence, ensuring uniform integrability of $\brac{\bar{\bm Q}^c(t)}_{c \in \Nats}$. 

Since the subsequence $\brac{\bar{\bm Z}^c(t),\bar{\bm Q}^c(t)}_{c \in \Nats}$ is uniformly integrable and $\brac{\bar{\bm Z}^c(T),\bar{\bm Q}^c(T)}$ converges in distribution to $(\bar{\bm Z}(T),\bar{\bm Q}(T)) = (\bm 0,\bm 0)$, we also have $L_1$ converges for the subsequence
\begin{equation}
\lim_{k\to \infty} \E \norm{\brac{\bar{\bm Z}^{c_k}(T),\bar{\bm Q}^{c_k}(T)}}_1=\E \norm{\brac{\bar{\bm Z}^{c_k}(T),\bar{\bm Q}^{c_k}(T)}}_1=0.
\end{equation}
The above $L_1$ convergence implies that there exists a $c_{\epsilon}$ such that for all $c>c_{\epsilon}$ we have
\begin{equation}
\label{eqn:ss_1}
\E \norm{\brac{\bar{\bm Z}^{c_k}(T),\bar{\bm Q}^{c_k}(T)}}_1 <(1-\delta),
\end{equation}
for any $\delta>0$.
Therefore, for $c=\norm{\brac{\bar{\bm Z}^{c_k}(T),\bar{\bm Q}^{c_k}(T)}}_1>c_\epsilon$ we can write 
\begin{equation}
\label{eqn:ss_2}
\E\sbrac{\norm{(\bm Z^c(cT) ,\bm Q^c\brac{cT})}_1-\norm{(\bm Z^c(0),\bm Q^c(0))}_1\mid \bm Q(0)}\leq -\delta c,
\end{equation}
where the inequality follows from~\eqref{eqn:ss_1}. The inequality in~\eqref{eqn:ss_2} shows that the conditions of the \textit{Multiplicative Foster's Lemma} are met [see Proposition 4.6 of~\cite{bramson2008stability}]. Therefore, the request queue process $\bar{\bm{Q}}^c$ under the policy $\pi^\star(\tau(c))$ is positive recurrent for all $c > c_\epsilon$. Thus we see that \eqref{eq:A} cannot hold. 
\hfill $\blacksquare$

\section{Azuma-Hoeffding Lemma}
We analyze a sequence of nested Martingale difference sequences with the following Azzuma-Hoeffding Lemma. Arguments of this type are regularly used for mixing bounds in reinforcement learning \cite{qu2020finite}. 

\begin{lem}
\label{lem:azzuma_hoeffding}
Let $$m_r(i,s)= \sbrac{\hat{n}^\star_r\brac{\bm Z(i\tau(c)+s),\bar{\bm Q}^c(i\tau(c)/c)}-\E \sbrac{\hat{n}^\star_r\brac{\bm Z(i\tau(c)+s),\bar{\bm Q}^c(i\tau(c)/c)} \mid \mathcal{F}_{i\tau(c)} }}, \ r\in \mathcal{R},$$ 
where $\mathcal{F}_{i\tau(c)} = \sigma\left(\bm{X}(s) : 0 \leq s \leq i\tau(c)\right)$ is the filtration\footnote{From the Markov property it implies that we can condition on filtration rather the process $\bm X$.} and $\left|m_r(i,s)\right|\leq B$ for all $i>0$, $s>0$. Then for any $\delta>0$ we have following bound
\begin{equation} \label{eq:lem1}
\mathbb{P}\brac{\sup_{u\leq t}\left| \frac{1}{c}\sum_{s=1}^{\tau(c)-1}\sum_{i:i\tau(c) \leq c u} m_r(i,s)\right|\geq \delta}\leq 2\tau(c) \exp{\frac{-c\delta^2}{2\tau(c) B^2 t}}.
\end{equation}
Thus for $\tau(c) = O(c^{1-\epsilon})$ for some $\epsilon$ we have, with probability 1, uniformly on compact time intervals
\[
\left| \frac{1}{c}\sum_{s=1}^{\tau(c)-1}\sum_{i:i\tau(c) \leq c u} m_r(i,s)\right|
\xrightarrow[c \rightarrow \infty]{} 0.
\]
\end{lem}

\begin{proof}
Using a union bound, we can write 
\begin{align*}
 \mathbb{P}\brac{\sup_{u\leq t}\left| \sum_{s=1}^{\tau(c)-1}\sum_{i:i\tau(c) \leq cu} m_r(i,s)\right|
 \geq c\delta}
 &
 \leq\mathbb{P}\brac{\exists s\leq \tau(c)-1:\sup_{u\leq t} \left|\sum_{i:i\tau(c) \leq c u} m_r(i,s)\right|\geq \frac{c\delta}{\tau(c)}}\\
 &
\leq \sum_{s=1}^{\tau(c)-1} \mathbb{P}\brac{\sup_{u\leq t}\left|\sum_{i:i\tau(c) \leq ct} m_r(i,s)\right|\geq \frac{c\delta}{\tau(c)}}.   
\end{align*}
 Note that $\sum_{i:i\tau(c) \leq ct} m_r(i,s)$ is a martingale difference sequence. Hence, the result follows by applying the Azuma-Hoeffding Inequality~\cite{hoeffding1994probability} to the last term in the above inequality. This can be applied to the maximum of the process by combining Azuma-Hoeffding with Doob's maximal inequality. 

For almost sure convergence, we note that the summation of the \eqref{eq:lem1} if finite for all $\delta>0$. By the Borel-Cantelli Lemma this implies almost convergence below $\delta$ and since this holds for all $\delta >0$ this implies almost sure convergence to zero as required.
\end{proof}

\section{Mixing Time Lemmas}\label{sec:mix}

We state two lemma on the coefficient of ergodicity. The proof of these can be found in the references  \cite{seneta1988perturbation,seneta2006non}.

The following lemma can be applied to show that the coefficient of ergodicity can be used to establish the mixing properties of a Markov chain.

 \begin{lem}\label{prop:ergprop}
\[
\rho(P P') \leq \rho(P)\rho(P')
\qquad
\text{and}
\qquad
|| \lambda P - \mu P ||_{\TV} \leq \rho(P) || \lambda - \mu ||_{\TV} \,.
\]
\end{lem}
The following lemma is due to Dobrushin \cite{dobrushin1956central}, and we apply this to prove Lemma \ref{lem:lemrho}.

\begin{lem}[Dobrushin's Lemma]
\begin{align*}
\rho(P)
&
=
1 -
\min_{z_1,z_2} \sum_{z \in \mathcal Z} \min \left\{ P_{z_1,z} , P_{z_2,z} \right\}	\, .
\end{align*}
\end{lem}

We can now prove Lemma \ref{lem:lemrho}, which ensures uniform fast mixing of LLEs.

\lemrho*
\begin{proof}
We apply Dobrushin's Lemma. Specifically, we note that for all $\bm Q$
\[
\mathbb P ( \bm Z(t+1) = \bm 0 | \bm Z(t) =\bm z , \bm Q(t) = \bm Q) \geq \delta^{B|\mathcal Z|}\, .
\] 
Note this corresponds to the event where the left-hand queue empties due to the decoherence of all LLEs.
Thus, it holds that
\[
\min_{z_1,z_2} \sum_{z \in \mathcal Z} \min \left\{ P^{\bm Q}_{z_1,z} , P^{\bm Q}_{z_2,z} \right\} \geq \min_{z_1,z_2}  \min \left\{ P^{\bm Q}_{z_1,\bm 0} , P^{\bm Q}_{z_2,\bm 0} \right\}  \geq \delta^{B|\mathcal Z|}\, .
\]
Thus, applying Dobrushin's Lemma:
\[
\rho(P^{\bm Q} )= 1-\min_{z_1,z_2} \sum_{z \in \mathcal Z} \min \left\{ P^{\bm Q}_{z_1,z} , P^{\bm Q}_{z_2,z} \right\} \leq 1-\delta^{B|\mathcal Z|} =: \rho <1\, ,
\]
as required.
\end{proof}




\section{Instablity Proof with Random Decoherence.}
\label{append:instability_maxweight}

We have already given a counter-example demonstrating that MaxWeight is not throughput optimal with deterministic LLE lifetimes and arrival processes. We now extend this argument for Bernoulli arrivals and fixed decoherence probabilities.

\MWInstable*

\begin{figure}[h]
    \centering
    \includegraphics[width=0.5\textwidth, trim=0cm 2cm 0cm 
    5cm, clip]{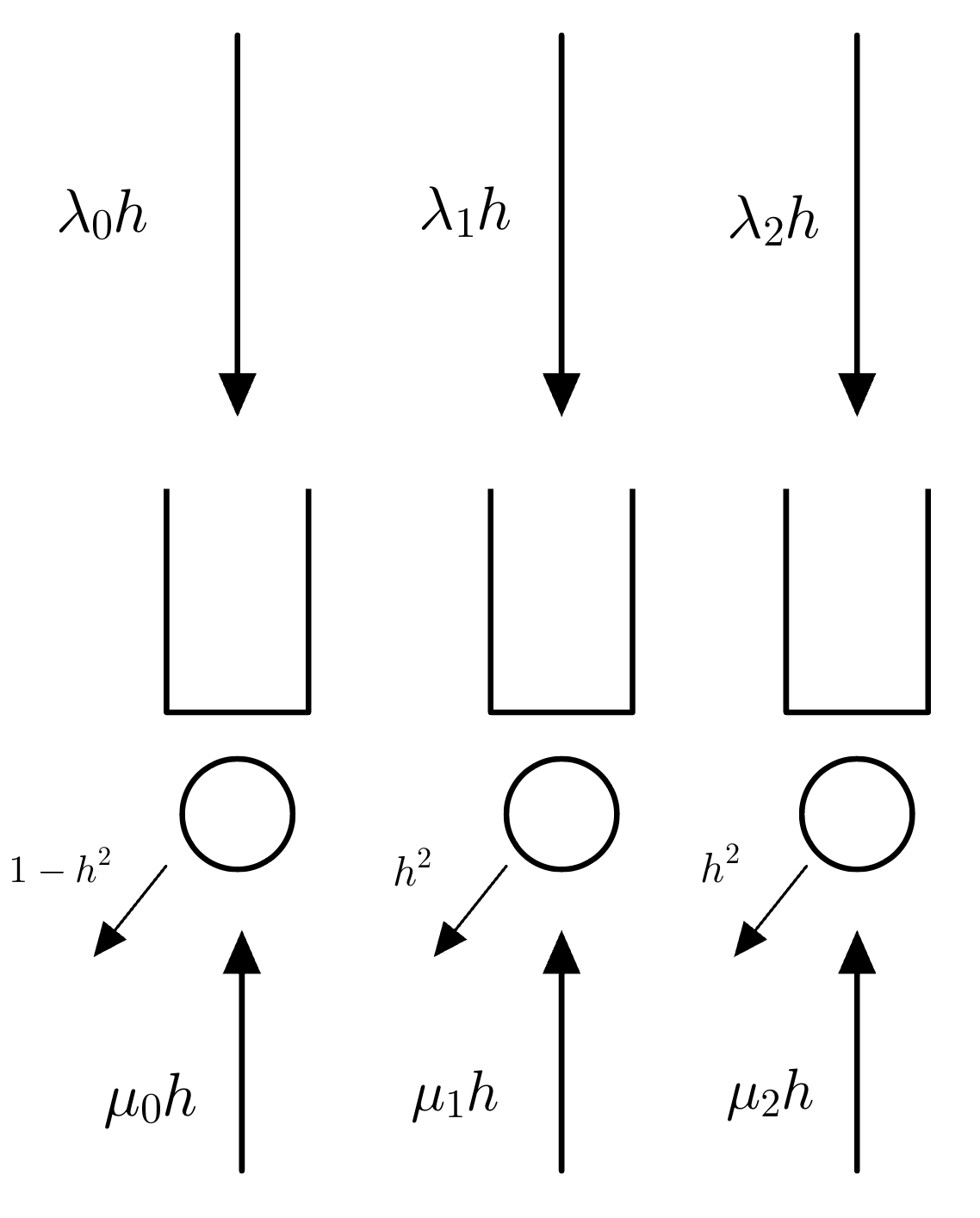}
    \caption{Queueing network topology with transitions}
    \label{fig:queue_network}
\end{figure}

\begin{proof}
\noindent \textit{Outline.} We develop a counter-example for the network described in Figure \ref{fig:queue_network}. Here, all transitions are Bernoulli random variables, unlike our previous counterexample.
In particular, request arrivals occur as Bernoulli processes with probabilities $\lambda_{i}h $ for $i = 0, 1, 2$. LLEs arrive with probabilities $\mu_i h$ for $i = 0, 1, 2$. In both cases, $h$ is a small but positive number.
Queue $0$ has a high probability of decoherence, $1-h^2$, whereas Queues $1$ and $2$ have a low probability of decoherence, $h^2$. For the network depicted in Figure~\ref{fig:queue_network}, type $0$ requests are served jointly by type $0$, type $1$, and type $2$ LLEs; type $1$ requests are served by type $1$ LLEs; and type $2$ requests are served by type-$2$ LLEs.

We consider two policies: MaxWeight and the policy that prioritizes Queue $0$, which we call P$0$. Find settings where the network is unstable under MaxWeight but stable under the P$0$ policy. Thus demonstrating that MaxWeight cannot be throughput optimal.

\medskip
\noindent \textit{A Continuous Time Limit.} To briefly give some intuition for the proof, notice if we let $h \rightarrow 0$ and we rescale time into intervals of length $h$, then our network converges to a continuous time Markov chain with arrival rates $\lambda_i$ and service rates $\mu_i$ for $i = 0, 1, 2$, where LLEs at Queues $1$ and $2$ have infinite lifetimes and the LLE at Queue $0$ decoheres instantaneously, unless it is served at the moment it arrives. Given the unavailability of Queue 0 LLEs, it is not too hard to show that for this continuous-time Markov chain, MaxWeight will prioritize Queues $1$ and $2$, whereas by definition, P$0$ prioritizes Queue 0. Thus, for MaxWeight, we only serve Queue 0 when Queues $1$ and $2$ are empty, which occurs at a proportion of $(1-\lambda_1/\mu_1)(1-\lambda_2/\mu_2)$. (Note Queues 1 and 2 behave independently under MaxWeight and are individual empty with probability $(1-\lambda_i/\mu_i)$, $i=1,2$.) Similarly, for the P$0$ policy, we only serve Queues $1$ and $2$ at a proportion of $(1-\lambda_0/\tilde \mu_0)$. 
(Here $\tilde \mu^{-1}_0$ is the mean service rate of Queue 0, which is the time for two LLEs to arrive at Queue $1$ and $2$ and for an LLE to arrive at Queue $0$. The exact formula for this is given below in (B).)
This leads to the following condition for the instability of MaxWeight and two conditions for the stability of the priority policy: 
\begin{align}
    \mu_0\bigg(1 - \frac{\lambda_1}{\mu_1}\bigg)\bigg(1 - \frac{\lambda_2}{\mu_2}\bigg) &< \lambda_0, \tag{A} \label{eqn:instability_maxweight} \\
    \frac{1}{\tilde{\mu}_0} := \frac{1}{\mu_1} + \frac{1}{\mu_2} - \frac{1}{\mu_1+\mu_2} + \frac{1}{\mu_0} &< \frac{1}{\lambda_0}, \tag{B}\label{eqn:p0stability1} \\
    \bigg(1 - \frac{\lambda_0}{\tilde{\mu}_0}\bigg) > \frac{\lambda_1}{\mu_1}\, , \quad 
    \text{and} 
    \quad \bigg(1 - \frac{\lambda_0}{\tilde{\mu}_0}\bigg) &> \frac{\lambda_2}{\mu_2}.
    \tag{C}\label{eqn:p0stability2}
\end{align}
The first condition~\eqref{eqn:instability_maxweight} is the condition for the instability of Queue $0$ under MaxWeight. The second condition~\eqref{eqn:p0stability1} is the condition for the stability of Queue 0 under the P$0$ policy. (Note that service requires waiting for two LLEs to arrive at Queues $1$ and $2$ and for an LLE to arrive at Queue $0$. The maximum of two exponential random variables plus an additional exponential has a mean of $	\tilde \mu^{-1}_0$ defined as above. Thus, the condition ensures the interarrival times are longer than service times.) The third condition~\eqref{eqn:p0stability2} is the condition for the stability of Queues $1$ and $2$ under the P$0$ policy. (Note that the long-run service rate of Queue $1$, which occurs when Queue $0$ is idle, is $\mu_1 (1-\lambda_0/\tilde{\mu}_0)$, and to be stable, this rate must be greater than $\lambda_1$.)

Thus, if we can find parameters for which~\eqref{eqn:instability_maxweight}-\eqref{eqn:p0stability2} are satisfied, we see that P$0$ is stable, and MaxWeight is not. Thus, MaxWeight cannot be optimal for throughput. This covers the case of the continuous-time model.
Some related proofs for CTMCs can be found in the paper of Bonald and Massouli\'e~\cite{bonald2001impact}. 
Now it remains to identify the parameters for which the conditions in~\eqref{eqn:instability_maxweight}-\eqref{eqn:p0stability2} are satisfied and to verify that the DTMC satisfies these conditions for sufficiently small $h$.

\medskip
\noindent \textit{Parameters Satisfying~\eqref{eqn:instability_maxweight}-\eqref{eqn:p0stability2}.} We verify that there exist parameters for which~\eqref{eqn:instability_maxweight}-\eqref{eqn:p0stability2} are satisfied. In particular, we can take

\[
\lambda_1 = \lambda_2 = 150, \quad \mu_1=\mu_2 = 200, \quad \lambda_0 = 4, \quad \mu_0 = 20 \, .
\]
For Condition~\eqref{eqn:instability_maxweight}:
\[
    \mu_0\bigg(1 - \frac{\lambda_1}{\mu_1}\bigg)\bigg(1 - \frac{\lambda_2}{\mu_2}\bigg)
    = 20 \times \frac{1}{4} \times \frac{1}{4} = 1.25 < 4 = \lambda_0 \, .
\]
For Condition~\eqref{eqn:p0stability1}:
\[
\frac{1}{\tilde{\mu}_0} := \frac{1}{200} + \frac{1}{200} - \frac{1}{400} + \frac{1}{20} = \frac{23}{400}  < \frac{1}{4} = \frac{1}{\lambda_0} \, .
\]
For Condition~\eqref{eqn:p0stability2}:
\[
\bigg(1 - \frac{\lambda_0}{\tilde{\mu}_0}\bigg) = 1 - \frac{92}{400} = \frac{308}{400} > \frac{3}{4} = \frac{\lambda_1}{\mu_1}= \frac{\lambda_2}{\mu_2} \, .
\]
Thus we see that there are parameters for which~\eqref{eqn:instability_maxweight}-\eqref{eqn:p0stability2} hold.

\medskip
\noindent \textit{Analysis for DTMC.} We have now completed the main conceptual components of the proof. What follows below is a detailed proof of the above inequalities, which accounts for the discrete transitions in the prelimit network, which depend on $h$. Here we show that~\eqref{eqn:instability_maxweight}-\eqref{eqn:p0stability2} hold for sufficiently small $h$. Moreover, we show that each expression is correct up to terms that are o(1) as $h \rightarrow 0$.\footnote{Such inequalities can likely be verified by a weak convergence argument. However, we provide direct proof here.}

Notice that excluding the transition for the decoherence of the type $0$ LLE, all transitions have a probability that is $O(h)$. Thus, the probability of a pair of any such transitions occurring simultaneously is $O(h^2)$. In particular, the probability of two transitions occurring simultaneously is bounded above by $h^2 C_0^{*}$ where $C_0^{*} = \max_{i=0,1,2}\{\mu^2_i, \lambda^2_i, 1\}$.  Thus, the probability of two transitions occurring simultaneously is $O(h^2)$ is bounded above by $ C_1^{*} = \binom{7}{2}  C_0^{*}$. Here, $\binom{7}{2}$ corresponds to the 28 pairs of transitions that can occur. We focus on verifying the instability condition for MaxWeight~\eqref{eqn:instability_maxweight} and then give conditions~\eqref{eqn:p0stability1} and~\eqref{eqn:p0stability2}.

\medskip
\noindent \textit{\eqref{eqn:instability_maxweight} Instability Condition for MaxWeight.}
We find parameters where Queue $0$ is starved out by the other two queues. We do this by lower bounding Queue $0$ and upper bounding Queues $1$ and $2$. In particular, we consider the lower-bound process where we assume every time there are two simultaneous transitions attempted, we remove a request from Queue $0$. 
This allows us to analyze situations where only one event occurs at a time since the worst that can happen from a pair of events is already factored into Queue $0$'s performance.

Now, notwithstanding the pairs of events just discussed, notice that whenever there is an LLE in Queue $1$ or $2$, that queue will be served if there is work to be done. In other words, unless two transitions occur simultaneously, Queues $1$ and $2$ have priority in effect over type $0$. Thus, excluding transitions with probability $O(h^2)$, Queue $0$ can only be served if both Queues $1$ and $2$ are empty.

\medskip
\noindent \textit{Idle Queue Conditions.}
Notice that for Queues $1$ (and Queue $2$), there is only one transition of order $O(h)$ that is influenced by other queues, 
that is, when the queue is empty, and there is service of an LLE waiting at Queue $1$. (This is because the LLE could be used to serve Queue $0$ or Queue $1$.)
This transition is always bounded below by $\lambda_1$, the probability of 
an arrival at Queue $1$. In other words, the transition at the queue and thus 
the queue size process at Queue $1$ can be bounded below by a single server 
queue with arrival probability $\lambda_1 h$ and departure probability $\mu_1 h - O(h^2)$. Thus we can bound above the time that the Queue $1$ spends idle with an LLE waiting by:
\[
P(\text{Queue $1$ is idle with an LLE}) \leq 1 - \frac{\lambda_1}{\mu_1} - O(h)\, .
\]
We can apply the same coupling to Queue $2$, and since the processes considered are independent
we have that
\[
P(\text{Both Queue 1 \& 2 idle with an LLE}) \leq \bigg(1 - \frac{\lambda_1}{\mu_1}\bigg)\bigg(1 - \frac{\lambda_2}{\mu_2}\bigg) + O(h).
\]
Thus, we must wait for an LLE to arrive. Thus, the rate of departure can be bounded above by:
\[
h \mu_0 \bigg(1 - \frac{\lambda_1}{\mu_1}\bigg)\bigg(1 - \frac{\lambda_2}{\mu_2}\bigg) + O(h^2) \, .
\]
Thus, with the above upper-bound on the long run departure rate from Queue $0$, we arrive at the condition that Queue $0$ will be unstable for $h$ sufficiently small, provided $\lambda$ and $\mu$ satisfy:
\[
    \mu_0 \bigg(1 - \frac{\lambda_1}{\mu_1}\bigg)\bigg(1 - \frac{\lambda_2}{\mu_2}\bigg) < \lambda_0 \, . \tag{A}
\]


\medskip
\noindent \textit{\eqref{eqn:p0stability1} Mean Service Time.} 
Conversely, suppose we prioritize Queue $0$. That is, we only use LLEs for Queues $1$ and $2$ when Queue $0$ is empty. In that case, 
let us bound the service rate at Queue $0$. Notice that since the lifetime 
of LLE's at Queue $1$ and $2$ are $O\left({h^{-2}}\right)$, thus the probability 
of the only way of serving Queue $0$ (with probability that is not $O(h^2)$) is 
that both LLEs at Queue $1$ and $2$ arrive (and do not depart), and then an LLE 
arrives at Queue $0$ and is served.

Note the LLE arrival at Queues $1$ and $2$ is the maximum of two geometric random variables, 
with mean $\frac{1}{h\mu_1} + \frac{1}{h\mu_2} - \frac{1}{h\mu_1+h\mu_2} + O(1)$, and the time for the LLE arrival at Queue $0$ is 
geometric with mean $\frac{1}{h\mu_0}$. Thus, the mean time between service 
epochs at Queue $0$ is:
\[
\frac{1}{h\mu_1} + \frac{1}{h\mu_2} - \frac{1}{h\mu_1+h\mu_2} + \frac{1}{h\mu_0} + O(1) \,.
\]
In other words, we have that the following condition is sufficient for stability:
\[
\frac{1}{\tilde \mu_0} := \frac{1}{\mu_1} + \frac{1}{\mu_2} - \frac{1}{\mu_1+\mu_2} + \frac{1}{\mu_0} < \frac{1}{\lambda_0} . \tag{B}
\]
That is, the mean time between service epochs at Queue $0$ is less than the mean time between arrivals at Queue $0$. This gives Condition~\eqref{eqn:p0stability1}.
(Notice that the service rate in (1) could be quicker if there are LLEs waiting in advance at Queues $0$ and $1$. Thus,~\eqref{eqn:p0stability1} is a pessimistic and sufficient condition for Queue $0$ to be stable.)

\medskip
\noindent \textit{\eqref{eqn:p0stability2} Sufficient Condition for Stability.} From the above service rate, we see that the time that the Queue $0$ spends idle is at least:
\[
\left(1 - \frac{\lambda_0}{\tilde \mu_0}\right) \, .
\]
Once idle, Queue 1 requests are served at rate $h \mu_1$. We have the following sufficient 
condition for stability:
\[
\mu_1 \left(1 - \frac{\lambda_0}{\tilde \mu_0}\right) > \lambda_1 \, ,
\]
or equivalently:
\[
\left(1 - \frac{\lambda_0}{\tilde{\mu}_0}\right) > \frac{\lambda_1}{ \mu_1}\, . \tag{C}
\]
This, along with the corresponding condition for Queue $2$, gives Condition~\eqref{eqn:p0stability2}.


\medskip
\noindent \textit{Conclusion.}
In summary, we see that conditions~\eqref{eqn:instability_maxweight}-\eqref{eqn:p0stability2} are satisfied for sufficiently small $h$. We have given parameters for which P$0$ can be stable but MaxWeight is not stable. Thus, MaxWeight is not optimal for throughput in the setting of a Quantum switch with entanglement memories.


\end{proof}

\begin{rem}
We note that by adding further queues, Queues $3, 4, 5, ..., N$, we can include further products in condition~\eqref{eqn:instability_maxweight}. This will further exacerbate MaxWeight's instability. This suggests that the loss of throughput can decrease multiplicatively with the network size.
\end{rem}

\end{document}